\PassOptionsToPackage{colorlinks,citecolor=blue,urlcolor=blue}{hyperref}
\documentclass[11pt]{article}

\usepackage[margin=1in]{geometry}
\usepackage{lipsum}
\usepackage{amsthm,amsmath,amsfonts,amssymb}
\usepackage[authoryear]{natbib}
\usepackage{xcolor}
\usepackage{float}
\usepackage{booktabs}
\usepackage{tikz}
\usepackage{orcidlink}
\usetikzlibrary{bayesnet}
\usetikzlibrary{positioning}
\usepackage[colorlinks,citecolor=blue,urlcolor=blue]{hyperref}
\usepackage{graphicx}
\usepackage{url}

\usepackage{float}
\usepackage{algorithm}
\usepackage{algpseudocode}
\usepackage{empheq}
\usepackage{empheq}
\usepackage{booktabs}   
\usepackage{multirow}   
\usepackage{etoolbox}
\usepackage{bbm}

\def\oper#1{\csdef{#1}{\operatorname{#1}}}
\forcsvlist\oper{GL,SL,deg,Hom,End,Vect,Mod,Rep,Ker,Im,Id}

\def\defbb#1{\csdef{b#1}{\mathbb{#1}}}
\forcsvlist\defbb{A,B,C,D,E,F,G,H,I,J,K,L,M,N,O,P,Q,R,S,T,U,V,W,X,Y,Z}

\def\defcal#1{\csdef{c#1}{\mathcal{#1}}}
\forcsvlist\defcal{A,B,C,D,E,F,G,H,I,J,K,L,M,N,O,P,Q,R,S,T,U,V,W,X,Y,Z}

\theoremstyle{plain}

\theoremstyle{definition}

\title{A sequential ensemble approach to epidemic modeling: Combining Hawkes and SEIR models using SMC$^2$}

\author{
  Dhorasso Temfack\,\orcidlink{0009-0007-0630-7504}\thanks{School of Computer Science and Statistics, Trinity College Dublin, Ireland. Email: \href{mailto:temfackd@tcd.ie}{temfackd@tcd.ie}} \and
 Jason Wyse\,\orcidlink{0000-0003-1391-7371}\footnotemark[1]
}

\date{} 

\begin{document}

\maketitle

\begin{abstract}
This paper proposes a sequential ensemble methodology for epidemic modeling that integrates discrete-time Hawkes processes (DTHP) and Susceptible-Exposed-Infectious-Removed (SEIR) models. Motivated by the need for accurate and reliable epidemic forecasts to inform timely public health interventions, we develop a flexible model averaging (MA) framework using Sequential Monte Carlo Squared. While generating estimates from each model individually, our approach dynamically assigns them weights based on their incrementally estimated marginal likelihoods, accounting for both model and parameter uncertainty, to produce a single ensemble estimate. We assess the methodology through simulation studies mimicking abrupt changes in epidemic dynamics, followed by an application to the  Irish influenza and COVID-19 epidemics. Our results show that combining the two models can improve both estimates of the infection trajectory and reproduction number compared to using either model alone. Moreover, the MA consistently produces more stable and informative estimates of the time-varying reproduction number, with credible intervals that provide a realistic assessment of uncertainty. These features are particularly useful when epidemic dynamics change rapidly, enabling more reliable short-term forecasts and timely public health decisions. This research contributes to pandemic preparedness by enhancing forecast reliability and supporting more informed public health responses.
\end{abstract}


\section{Introduction}

Emerging infectious diseases pose significant challenges to public health systems worldwide. In situations where no effective vaccine is available, governments must rapidly implement a range of non-pharmaceutical interventions (NPIs) to mitigate the spread of the virus. From social distancing measures to intensive lockdowns, these strategies are often guided by mathematical and statistical models \citep{hellewell2020feasibility, fowler2020effect}. Statistical models enable policymakers to simulate multiple scenarios and quantify the efficacy of interventions, thus informing decisions aimed at minimizing the burden on healthcare systems. 


Among the most widely adopted frameworks are compartmental models, such as the SIR (Susceptible-Infectious-Removed) or SEIR (Susceptible-Exposed-Infectious-Removed) model, which provide an intuitive representation of disease dynamics by tracking transitions between health states \citep{kermack1927contribution}. Compartmental models are typically described by a system of equations that govern the rates at which individuals move between compartments based on disease-specific parameters. A widely used approach for estimating the parameters of such models is Bayesian inference, which provides a principled way to quantify uncertainty in model estimates and incorporate prior knowledge. SEIR-type models have been extensively applied to various infectious diseases, including Ebola \citep{lekone2006statistical, funk2018real}, Influenza \citep{Dureau2013, birrell2020efficient}, and COVID-19 \citep{cazelles2021mechanistic, storvik2023sequential}. Their modular structure also makes them easily extendable to incorporate additional compartments or features, such as vaccination, age structure, or spatial dynamics. Moreover, they enable the estimation of key epidemiological quantities, such as the incubation and infectious periods, which are directly interpretable by epidemiologists. However, increasing model complexity can introduce challenges, including parameter identifiability issues, potential biases, and susceptibility to mis-specification, all of which require careful consideration during model development and inference \citep{kresin2021comparison}. In response, researchers have explored alternative modeling frameworks, including point process models such as the Hawkes process.

The Hawkes process \citep{hawkes1971point} is a type of self-exciting point process where past events influence the likelihood of future occurrences. This model has gained attention for its ability to capture event clustering and self-excitation, making it particularly suitable for modeling infectious disease case series. A notable early use of point process models for infectious diseases, though not Hawkes-based, was by \cite{meyer2014power}. An early Hawkes-based application to real epidemic data was proposed by \cite{kelly2019real}, who developed a nonparametric model to forecast the 2018–2019 Ebola outbreak in the Democratic Republic of the Congo. During the COVID-19 pandemic, Hawkes processes were employed to estimate the time-evolution of the reproduction number \citep{mohler2021analyzing, chiang2022hawkes, lamprinakou2023using}. Several studies have compared Hawkes processes with compartmental models. \cite{bertozzi2020challenges} applied a continuous-time Hawkes model to COVID-19 data in U.S. states, but it underperformed relative to traditional SIR/SEIR models during early epidemic phases. In contrast, \cite{kresin2021comparison} reported that the Hawkes process outperformed SEIR, reducing forecast error by 20–30\% in most cases. \cite{park2022non} found that a Hawkes model improved forecasts by 38\% over SEIR during the 2014–2016 Ebola outbreak. Furthermore, \cite{rizoiu2018sir} introduced the HawkesN process, bridging Hawkes and SIR frameworks by incorporating finite population constraints. They showed that, in the absence of background events and assuming an exponential infectious period, the expected conditional intensity of HawkesN matches that of the stochastic SIR model, providing a mechanistic link between these approaches.

Unlike compartmental models, a Hawkes process models infections using a single conditional intensity function, avoiding assumptions about unobserved states. However, in real-world scenarios, the exact timing of each infection is rarely recorded. Instead, data are reported in aggregated time intervals (e.g., daily or weekly). To address this, the discrete-time Hawkes process (DTHP) was developed to accommodate aggregate case data without requiring artificial interpolation of infection times. This approach has been applied in various contexts, including modeling the spread of COVID-19 \citep{browning2021simple, koyama2021estimating}.

Having access to appropriate modeling and prediction techniques during an emerging epidemic is critical, as their outputs can play a prominent role in policy decisions and intervention strategies. A common approach is to assume the data follows one specific model. To address this limitation, we propose an ensemble modeling framework that integrates estimates from both the DTHP and a stochastic SEIR model, accounting for uncertainty in both model structure and parameters. We employ the Sequential Monte Carlo Squared (SMC$^2$) framework of \cite{Chopin2013}, adapted to jointly estimate the latent states and parameters of each model. By weighting model outputs based on their predictive performance, we dynamically combine their estimates. Our approach focuses on online inference of the case incidence and the instantaneous reproduction across models.

 While Bayesian model averaging is well-established in infectious disease modeling \citep{castaneda2015accounting, park2017ensemble, goudie2019joining}, most existing approaches rely on Markov Chain Monte Carlo (MCMC) methods to compute posterior model probabilities. These require access to the full dataset, rendering them impractical for real-time applications. To address this, Bayesian model averaging using SMC methods has been proposed to sequentially integrate predictions from multiple models, weighted by their posterior likelihoods of generating observed data \citep{parrish2012toward, rings2012bayesian, martino2017cooperative, urteaga2016sequential}, though these approaches have not yet been applied to epidemiological modeling. Moreover, they typically assume known model parameters, neglecting internal parameter uncertainty, a condition rarely met in practice. Our work builds on and extends these efforts by combining independent SMC$^2$ inference for each model with sequential Bayesian model averaging. Specifically, the DTHP and SEIR models are run simultaneously to infer states and parameters. At each step, model-specific posterior predictive distributions are combined using posterior model probabilities computed from the SMC$^2$-estimated model evidence. This procedure does not fully embed BMA within the SMC$^2$ algorithm; instead, model averaging is applied on the dynamic outputs of the independent SMC$^2$ updates. This allows real-time updating of predictions while accounting for uncertainty in both the model structure and the parameters. A valuable byproduct of this approach is the ability to estimate key epidemiological quantities, such as the infectious period from the SEIR model, and to assess how past events influence current transmission dynamics via the DTHP triggering kernel, providing complementary insights alongside primary predictions.

The remainder of this paper is organized as follows. In Section \ref{sec2}, we introduce the DTHP and stochastic SEIR models used in this study. Section \ref{sec3} presents the Bayesian model averaging framework and the sequential estimation procedure via SMC$^2$. Section \ref{sec4} details the experimental setup and the analysis of real data from the Irish influenza and COVID-19 epidemics. Finally, Section \ref{sec5} concludes with a summary of our findings and their implications for future research. The source code for this study is publicly available at \href{https://github.com/Dhorasso/bma-smc2-dthp-seir}{https://github.com/Dhorasso/bma-smc2-dthp-seir}.


\section{Statistical models}\label{sec2}

A central objective of this work is to develop a unified Bayesian framework that allows complementary epidemic models to be merged to carry out real-time inference. To this end, we consider the DTHP  and a stochastic SEIR model. These models provide distinct yet compatible representations of transmission dynamics, self-excitation in the DTHP, and mechanistic compartmental structure in the SEIR model. Both models incorporate a shared quantity of interest, such as the time-varying reproduction number $R_t$, which quantifies the average number of secondary infections generated by a single infectious individual at time $t$. While estimating $R_t$ is not the main focus of our framework, combining the DTHP and SEIR models within a single Bayesian procedure yields a unified estimate of $R_t$, providing a meaningful and interpretable measure of transmissibility alongside other outputs.

\subsection{Discrete-Time Hawkes Process}\label{sec_dthp}
The Discrete-Time Hawkes Process (DTHP) is a stochastic point process specifically designed to model the occurrence of events such as new cases of an infectious disease over discrete time intervals \citep{koyama2021estimating,browning2021simple}. It extends the continuous-time Hawkes process to settings where event times are unobserved and data are reported as aggregated counts over fixed intervals. Mathematically, a DTHP model is characterized by a conditional intensity function $\lambda_{H}(t)$, which represents the expected number of cases occurring at time $t$, given the history of previous cases. Unlike the continuous-time formulation, where the event history $\mathcal{H}_{t-1}$ typically consists of precise event times and associated marks, the discrete-time version uses only aggregated past counts. Specifically, the history is given by $\mathcal{H}_{t-1} := \{y_s| 1 \leq s \leq t - 1\}$, where $y_s$ denotes the number of cases observed at time $s$. The model is defined as follows:
\begin{align}\label{dthp}
\mathcal{M}_1:\begin{dcases}
\lambda_{H}(t) = \left(1 - \dfrac{\sum_{s=1}^{t-1} y_s}{N}\right) \left[\mu + R_t \sum_{s=1}^{t-1} y_s \varphi_{t-s}\right], \\
\log(R_t) = \log(R_{t-1}) + \varepsilon_{1,t}, \quad \varepsilon_{1,t} \sim \mathcal{N}(0, \nu^2_{1})
\end{dcases}
\end{align}
Here $\mu$ represents the expected number of spontaneous or imported cases. The term $\sum_{s=1}^{t-1} y_s \varphi_{t-s}$ captures the cumulative self-excitation process, the mechanism by which $y_s$ cases that occurred at time $s$ exert influence with a delay of $ t-s$ units of time. The function $\varphi_{t-s}$ is the triggering kernel, describing the discrete distribution of the delay $d=t-s$ between infection and subsequent transmission, and satisfies $\sum_{d=1}^{\infty} \varphi_{d} = 1$. We adopt a geometric kernel, defined as $\varphi_d = \omega (1 - \omega)^{d - 1}$. This choice is motivated by the relationship between the geometric and exponential distributions: in continuous-time Hawkes processes, the exponential kernel is widely used to model memory decay. While this formulation assumes maximal influence immediately after infection, it can be adapted for diseases with a latent period by choosing a kernel with a mode at $d>1$. The geometric kernel provides a simple yet flexible starting point for general epidemics and is a natural discrete-time counterpart to the exponential distribution, capturing how infectiousness decays over time \citep{browning2021simple}. In this context, $\omega\in(0,1)$ controls the rate at which the influence of past cases diminishes over time. The factor $\left(1 - \frac{\sum_{s=1}^{t-1} y_s}{N}\right)$ accounts for the reduction in the susceptible population due to prior infections, ensuring that the model reflects the decreasing pool of individuals at risk as the epidemic progresses; $N$ denotes the total population size. Since the true trajectory of the reproduction number $R_t$ is unknown and likely to change over time due to interventions or behavioral shifts, we model $\log(R_t)$ as a random walk. This flexible formulation allows $R_t$ to adapt dynamically to change and seems more suited than a stationary process, which implies time-invariant statistical properties. The log-transform ensures positivity of $R_t$, a biologically meaningful constraint. This approach is widely adopted in both continuous and discrete-time epidemic models \citep{abbott2020estimating, koyama2021estimating, lamprinakou2023using}. To promote numerical stability, we place a prior on the innovation variance $\nu_1$ that imposes a penalty against excessive variability in $\log(R_t)$, thereby limiting unrealistically rapid fluctuations or runaway growth in $R_t$. At time $t=0$, latent variables are assigned values from prior distributions, which encode uncertainty about the initial state of the epidemic.

\subsection{Stochastic SEIR model}
The stochastic SEIR model provides a framework for characterizing transitions between different stages of an epidemic. Susceptible individuals become exposed upon contact with infectious individuals. Exposed individuals transition to the infectious state after an incubation delay, and infectious individuals are eventually removed through recovery or death. The number of individuals transitioning between compartments are modeled as binomial random variables. The model is defined by the following update equations, based on \cite{lekone2006statistical}:
\begin{align}\label{seir}
\mathcal{M}_2:\begin{dcases}
S(t+1) = S(t) - \lambda_{SE}(t), \hspace{1.8cm} \lambda_{SE}(t) \sim \text{Binomial}\left(S(t), 1-e^{-\beta_t \frac{ I(t)}{N}}\right) , \\
E(t+1) = E(t) + \lambda_{SE}(t) - \lambda_{EI}(t), \hspace{0.4cm} \lambda_{EI}(t) \sim \text{Binomial}\left(E(t), 1-e^{-\sigma}\right), \\
I(t+1) = I(t) + \lambda_{EI}(t) - \lambda_{IR}(t), \hspace{0.5cm} \lambda_{IR}(t) \sim \text{Binomial}\left(I(t), 1-e^{-\gamma }\right), \\
R(t+1) = R(t) + \lambda_{IR}(t), \\
\log(\beta_t) = \log(\beta_{t-1}) + \varepsilon_{2,t}, \hspace{1.8cm} \varepsilon_{2,t} \sim \mathcal{N}(0, \nu^2_{2}), \\
R_t = \dfrac{\beta_t}{\gamma}.
\end{dcases}
\end{align}
In this model, $S(t)$, $E(t)$, $I(t)$, and $R(t)$  represent the numbers of susceptible, exposed  (infected but
not yet infectious), infectious, and removed (recovered or dead) individuals at time $t$, respectively. We assume a closed population, i.e. $N = S(t) + E(t) + I(t) + R(t)$ for all $t \in \mathbb{N}$. While the classic SEIR model with fixed parameters provides a useful baseline for understanding epidemic dynamics, it is structurally limited in its ability to capture complex phenomena such as multiple waves of infection. To overcome this, we augment the model with a time-varying transmission rate $\beta_t$, which evolves according to a log-normal random walk, analogous to the modeling of $R_t$ in DTHP discussed in Section \ref{sec_dthp}. The parameters $1/\sigma$ and $1/\gamma$ represent the average durations of the latent and infectious periods, respectively. The quantities $\lambda_{SE}(t)$, $\lambda_{EI}(t)$, and $\lambda_{IR}(t)$ represent new exposures, new infections, and recoveries, while $R_t$ denotes the time-varying reproduction number. The initial values for latent variables at time $t=0$, are drawn from prior distributions chosen to reflect uncertainty in the epidemic's starting state.

Both the DTHP and SEIR models incorporate a random walk (RW) on $R_t$ in the DTHP and $\beta_t$ in the SEIR model, allowing flexible adaptation to changing transmission dynamics. As a simplifying modeling choice, the remaining parameters are assumed approximately constant over the epidemic timescale. Unlike transmission, which can respond quickly to interventions or behavioral shifts, parameters such as the infectious or incubation period typically vary slowly and are not expected to exhibit abrupt changes. While fully dynamic parameter models \citep{liu_west_2001, lamprinakou2023using} could in principle be applied, they can increase variance in likelihood estimates, risk filter degeneracy, and, as noted in \cite{Chopin2013}, produce systematically biased model evidence. Consequently, we focus on temporal flexibility on transmission-related parameters while treating other epidemiological parameters as effectively static. The magnitude of RW fluctuations is controlled by the volatility parameters $\nu_k$ ($k=1,2$), which are sequentially estimated from observed incidence data via the SMC$^2$ algorithm (Section \ref{sec3}); higher $\nu_k$ values correspond to larger changes in transmission. As shown in Section \ref{sec4}, this targeted approach enables accurate tracking of the reproduction number using observed incidence data. Other extensions of the SEIR model are discussed later.

\subsection{Observation process}
Let $y_t$ denote the number of observed cases on day $t = 1, \ldots, T$. For each model $\mathcal{M}_k$, we denote by $\lambda_{k,t}$ the expected number of new infections at time $t$, defined as
\begin{align}
    \lambda_{k,t} = 
    \begin{cases}
        \lambda_{H}(t), & \text{DTHP model if } k = 1, \\
        \lambda_{EI}(t), & \text{SEIR model if } k = 2.
    \end{cases}
\end{align}
During an epidemic, daily reported case counts are subject to various sources of noise, including measurement error, behavioral factors, and inherent randomness. To account for this, we model the observed case counts using a negative binomial distribution \citep{koyama2021estimating, inouzhe2023}, with mean $\lambda_{k,t}$ and variance $\lambda_{k,t} + \phi_k \lambda^{2}_{k,t}$, where $\phi_k > 0$ is an overdispersion parameter, i.e.,
\begin{align}\label{ob_proces}
    y_t |  \lambda_{k,t} \sim \text{NegBin}(\lambda_{k,t},\, \lambda_{k,t} + \phi_k \lambda^{2}_{k,t}), \quad k \in \{1, 2\},
\end{align}
 with a probability mass function
\begin{align}\label{ydist_pmf}
    g(y_t| \lambda_{k,t}, \phi_k) = \dfrac{\Gamma\left(y_t + \frac{1}{\phi_k}\right)}{y_t!\Gamma\left(\frac{1}{\phi_k}\right) }\left(\dfrac{1}{1+\phi_k\lambda_{k,t}}\right)^ {\frac{1}{\phi_k}}\left(\dfrac{\phi_k\lambda_{k,t}}{\phi_k\lambda_{k,t}+1}\right)^ {y_t}, \quad k\in\{1,2\}.
\end{align}
As $\phi_k$ tends to $0$, the variance approaches the mean, and we have convergence to a Poisson probability mass function. The negative binomial distribution is widely used to accommodate overdispersed count data, providing a more flexible fit than the Poisson distribution in such contexts \citep{linden2011using}. Importantly, this formulation assumes perfect case detection and does not explicitly account for case ascertainment fractions or reporting delays. However, for more practical applications, such factors can easily be incorporated into the model through additional latent processes or delay distributions \citep{Birrell2017, abbott2020estimating}. Although $\lambda_{1,t}$ and $\lambda_{2,t}$ model the same expected incidence at time $t$, we allow model–specific overdispersion parameters $\phi_1 \neq \phi_2$. The two models generate incidence through different latent mechanisms, which leads to different levels of predictive variability. Using separate $\phi_k$ therefore provides each model with an appropriate observation noise structure and also allows the two SMC$^2$ algorithms to be run independently. Similar multi-model frameworks routinely employ model-specific observation noise distributions \citep{park2017ensemble, adiga2021all}.


\section{Sequential inference and Bayesian model averaging }\label{sec3}

SMC-based methods facilitate approximate inference in Bayesian sequential models by representing the evolving posterior over latent states through a set of weighted particles, making them well-suited for real-time data assimilation. Building on this, model averaging combines predictions from multiple candidate models, weighting them by their posterior model probabilities obtained via the SMC framework \citep{martino2017cooperative, urteaga2016sequential}. This approach incorporates model uncertainty into the analysis, resulting in more robust and reliable predictions.

\subsection{State-space formulation}
Assume the observed data are generated by one of a set of candidate models $\{\mathcal{M}_k\}_{k=1}^{K}$, where $K \geq 2$. In this study, we focus on the case $K=2$, corresponding to the DTHP and SEIR models, but the formulation below is presented for a general number of models. Following \citet{jacob2017bayesian}, each model $\mathcal{M}_k$ is specified through a state-space representation at discrete times $t = 1, \ldots, T$:
\begin{align}\label{ssmk}  
    & x_{k,0} \sim f_{k}(x_{k,0} |\theta_{k}),  \hspace{2.65cm} & \rhd & \quad \text{Initial state}\\
     & x_{k,t}  | x_{k, 0:t-1}\sim f_{k}(x_{k,t} | x_{k, 0:t-1}, \theta_{k}),    \hspace{0.5cm} &\rhd &  \quad  \text{State process}   \\
    & y_{t}| x_{k,t} \sim g(y_{t} | x_{k,t}, \theta_{k}),     \hspace{2cm}    &\rhd &\quad   \text{Observation process}
\end{align}
where the latent state $x_{k,t} \in \mathbb{R}^{n_{x_k}}$ is propagated according to the transition density $f_{k}(x_{k,t} | x_{k, 0:t-1}, \theta_{k})$, which is simply one of the statistical models given in Equations \eqref{dthp} and \eqref{seir}. The observations $y_t \in \mathbb{R}$ are linked to the latent states through the emission density $g(y_{t} | x_{k,t}, \theta_{k})$, as in Equation \eqref{ydist_pmf} and the vectors $\theta_{k}\in \Theta_{k} \subset \mathbb{R}^{n_{\theta_k}}$ represent static parameters in the corresponding model. For any sequence $\{z_t\}_{t\geq 0}$, we denote $z_{i:j} = \{z_i, z_{i+1}, \ldots, z_j\}$.

When $k = 1$, the state-space model corresponds to the DTHP model, as given in \eqref{dthp}, with the latent state $x_{1,t} = \left( \lambda_{H}(t), R_t \right)^{\top}$ and parameters $\theta_{1} = \left( \mu, \omega, \nu_1, \phi_1 \right)^{\top}$.

For $k = 2$, it corresponds to the SEIR model, as given in \eqref{seir}, with the latent state $x_{2,t} = (S(t), E(t), I(t), R(t), \lambda_{EI}(t), R_t)^{\top}$ and parameters $\theta_{2} = \left( \sigma, \gamma, \nu_2, \phi_2 \right)^{\top}$.

For a given value of $\theta_{k}$, inference is performed on the sequence of latent variables $x_{k,t}$ given the observations $y_{1:t},~t\geq 1$ using the conditional posterior distribution $p(x_{k, t} | y_{1:t}, \theta_{k}, \mathcal{M}_k)$. Bayesian model averaging then provides a formal probabilistic framework to obtain predictive inference on quantities of interest that are not model-specific \citep{Hoeting99}. That is the posterior distribution of a shared quantity of interest at time $t$, denoted $\Delta_t$, which has a consistent interpretation across all considered models (e.g., the expected number of newly infected cases or the time-varying reproduction number $R_t$), can be expressed as a weighted mixture of the model-specific posterior distributions.

To account for uncertainty in the model parameters, we integrate over the posterior distribution of $\theta_k$. Denoting the posterior over parameters given model $\mathcal{M}_k$ and observations $y_{1:t}$ as $p(\theta_k|y_{1:t}, \mathcal{M}_k)$, the model-specific posterior predictive distribution for a quantity of interest $\Delta_t$ is obtained by marginalization:
\begin{align}\label{marg_predictive}
p(\Delta_t | y_{1:t}, \mathcal{M}_k) = \int p(\Delta_t | y_{1:t}, \theta_k, \mathcal{M}_k) \, p(\theta_k | y_{1:t}, \mathcal{M}_k) \, d\theta_k.
\end{align}
Bayesian model averaging then combines these model-specific predictive distributions to produce a unified posterior for $\Delta_t$:
\begin{align}\label{bma_marginal}
p(\Delta_t | y_{1:t}) = \sum_{k=1}^{K} \pi_{k,t}\, p(\Delta_t | y_{1:t}, \mathcal{M}_k),
\end{align}
where the weights $\pi_{k,t}$ are the posterior model probabilities:
\begin{align}\label{true_pi}
\pi_{k,t} = \dfrac{p(y_{1:t} | \mathcal{M}_k)\, p(\mathcal{M}_k)}
{\sum_{j=1}^{K} p(y_{1:t}| \mathcal{M}_j)\, p(\mathcal{M}_j)},
\quad k = 1,\ldots,K.
\end{align}
The model evidence $p(y_{1:t} | \mathcal{M}_k)$, is obtained by integrating over the model parameters:
\begin{align}\label{marg_evidence}
p(y_{1:t}| \mathcal{M}_k)
&= \prod_{s=1}^{t} p(y_s|y_{1:s-1}, \mathcal{M}_k) \notag \\
&= \prod_{s=1}^{t} \int p(y_s | y_{1:s-1}, \theta_k, \mathcal{M}_k)\,
p(\theta_k | y_{1:s-1}, \mathcal{M}_k)\, d\theta_k,
\quad k = 1,\ldots,K.
\end{align}
Here, $p(y_s \mid y_{1:s-1}, \theta_k, \mathcal{M}_k)$ denotes the incremental likelihood at time $s$. In this study, equal prior probabilities are assigned to all models, $p(\mathcal{M}_k) = 1/K$ for $k=1,\dots,K$, reflecting no a priori preference among the candidate models.

\subsection{Sequential Monte Carlo for known parameters}

For clarity, we present the particle filtering algorithm without the model index $k$, as all operations are performed independently for each model. When the model parameters $\theta$ are known, Sequential Monte Carlo (SMC), also referred to as particle filtering \citep{gordon1993novel, doucet2001introduction}, recursively approximates the filtering distribution of the latent variables at each time point, given the data up to time $t$. This is achieved by sequentially generating a set of particles with associated weights, denoted by $\{x_t^i, w_t^i\}_{i=1}^{N_x}$, using importance sampling and resampling techniques. Here, $N_x$ denotes the number of particles, and $w_t^i$ represents the unnormalized importance weight of the $i$-th particle at time $t$, defined as:
\begin{align} \label{wt2}
    w_t^i = w_{t-1}^i \dfrac{g(y_t| x_t^i, \theta) \, f(x_t^i| x_{0:t-1}^i,\theta)}{q(x_t^i| x_{0:t-1}^i, y_t, \theta)}, \quad i = 1, \ldots, N_x,
\end{align}
where $x_t^i$ is sampled from the importance distribution $q(.| x_{0:t-1}^i, y_t, \theta)$. At each time $t$, the filtering distribution of the latent state can then be approximated by a weighted sum of Dirac delta functions:
\begin{align}
   p(x_t | y_{1:t}, \theta) \approx \sum_{i=1}^{N_x} W_t^i \, \delta_{x_t^i}(x_t), \qquad W_t^i = w_t^i / \sum_{j=1}^{N_x} w_t^j.
\end{align}
After the resampling step, the posterior distribution is approximated as:
\begin{align}
   p(x_t | y_{1:t}, \theta) \approx \dfrac{1}{N_x} \sum_{i=1}^{N_x} \delta_{x_t^{a_t^i}}(x_t),
\end{align}
where $a_t^i$ denotes the resampled ancestor index at time $t$ (see Algorithm \ref{BMA-pf}).

A notable particle filtering method employed in this study is the Bootstrap Particle Filter (BPF) \citep{gordon1993novel}, which simplifies weight calculations by setting the importance distribution equal to the state transition distribution, i.e., $q(x_t \mid x_{0:t-1}, y_t, \theta) = f(x_t \mid x_{0:t-1}, \theta)$. The particle filter propagates multiple realizations (particles) of the system state forward in time using Equation  \eqref{dthp} for the DTHP  and Equation \eqref{seir} for the SEIR model, with each particle assigned a weight proportional to the likelihood of its simulated observations matching the actual data. A detailed description of the BPF procedure is provided in Algorithm~\ref{BMA-pf}.

\begin{algorithm}[H]
\caption{Bootstrap Particle Filter (BPF) for known parameters} \label{BMA-pf}
\begin{flushleft}
Operations involving index $i$ must be performed for $i = 1,\dots, N_x$.

The indices $a_{t}^{1:N_x}$ define the ancestral state particles at time $t$ after resampling.

\textbf{Inputs:} Observations $y_{1:T}$, number of particles $N_x$, model parameters $\theta$.

\textbf{Outputs:} Particle set $\{x_{0:t}^i, w_{0:t}^i\}_{i=1}^{N_x}$, marginal likelihood estimate $\widehat{p}_{_{N_x}}(y_{1:t}|\theta)$.
\end{flushleft}

\hrulefill 
\begin{algorithmic}[1]
    \State Sample initial particles: $x_0^i \sim f(x_0 | \theta)$
    \State Initialize weights: $w_0^i = 1$, $W_0^i = 1/N_x$
    \For{$t=1$ to $T$}
        \State \label{res1} Sample ancestor indices: $a_t^{1:N_x} \sim \text{Resample}(W_{t-1}^{1:N_x})$
        \State \label{fk}Propagate states: $x_t^i \sim f(\cdot | x_{0:t-1}^{a_t^i}, \theta)$
        \State \label{wpmf} Compute and normalize weights: $w_t^i = g(y_t | x_t^i, \theta), \quad W_t^i = w_t^i / \sum_{j=1}^{N_x} w_t^j$
        \State Compute marginal likelihood estimate: $\widehat{p}_{_{N_x}}(y_{1:t}|\theta) = \prod_{s=1}^t \left(\frac{1}{N_x} \sum_{i=1}^{N_x} w_s^i\right)$
    \EndFor
\end{algorithmic}
\end{algorithm}

The resampling step, outlined in Step \ref{res1} of Algorithm \ref{BMA-pf}, eliminates particles with low weights and replicates those with higher weights. After resampling, all particle weights are reset to one. Unless stated otherwise, we employ stratified resampling due to its efficiency and lower computational cost \citep{douc2005comparison, vieira2018bayesian}. Step \ref{fk} involves applying the transition density $f$, corresponding to the state update rule of the chosen model (DTHP or SEIR). Step \ref{wpmf} entails evaluating the probability mass function of the observation model (Equation \eqref{ydist_pmf}). In cases with missing observations, we set $w_t^i = 1$, for $i=1, \dots, N_x$.  A notable feature of particle filtering is that it provides an unbiased estimate of the incremental likelihood. Specifically, we have
\begin{align}\label{int_inc}
p(y_t | y_{1:t-1}, \theta) = \int g(y_t| x_t, \theta) \, p(x_t | y_{1:t-1}, \theta) \, dx_t
= \mathrm{E}_{\,p(x_t | y_{1:t-1}, \theta)}[g(y_t| x_t, \theta)],
\end{align}
and noting that, in the case of the BPF, the weights are exactly $w_t = g(y_t | x_{t}, \theta)$, we can use the trajectories $\{ x^{i}_{k,t} \}_{i=1}^{N_x}$ sampled from the one-step-ahead predictive density $p(x_{t} | y_{1:t-1}, \theta)$ and use a Monte Carlo approximation of the expectation to obtain:
\begin{align}\label{inc_bpf}
\widehat{p}_{_{N_x}}(y_t | y_{1:t-1}, \theta) = \dfrac{1}{N_x} \sum_{i=1}^{N_x} w_t^i.
\end{align}
Hence, the marginal likelihood up to time $t$ for a given $\theta$ is approximated by the product of incremental likelihoods:
\begin{align}\label{malik}
\widehat{p}_{_{N_x}}(y_{1:t} | \theta) = \prod_{s=1}^t \widehat{p}_{_{N_x}}(y_s | y_{1:s-1}, \theta) = \prod_{s=1}^t \left( \dfrac{1}{N_x} \sum_{i=1}^{N_x} w_s^i \right).
\end{align}
In many realistic situations, the true value of the parameter $\theta$ is unknown. While expert knowledge can sometimes be used to inform epidemiological quantities, such as the average infectious period, it is more challenging to specify suitable values for hyperparameters governing the innovations of the Brownian process and the observation distribution. To carry out Bayesian learning of the parameters from the observations, expert knowledge is encoded in a prior probability distribution $p(\theta)$, and the aim is to estimate the sequence of posterior distributions given the observed data.

\subsection{Bayesian model averaging via SMC$^2$} \label{sec:smc2}
SMC$^2$ targets the sequence of posterior distributions over parameters, $\{p(\theta_{k} | y_{1:t}, \mathcal{M}_k), ~ t=1,\dots, T\}$ which evolve as new data becomes available. Given a prior density $p(\theta_k)$ over the parameters $\theta_k$, Bayes' Theorem gives: 
\begin{align}\label{baye}
p(\theta_{k} | y_{1:t}, \mathcal{M}_k)
&= \dfrac{p(y_t | y_{1:t-1}, \theta_{k}, \mathcal{M}_k)\, p(\theta_{k} | y_{1:t-1}, \mathcal{M}_k)}
{\int p(y_t | y_{1:t-1}, \theta_{k}, \mathcal{M}_k)\, p(\theta_{k} | y_{1:t-1}, \mathcal{M}_k)\, d\theta_k} \notag\\
&\propto p(y_t | y_{1:t-1}, \theta_{k}, \mathcal{M}_k) \, p(\theta_{k} | y_{1:t-1}, \mathcal{M}_k),
\end{align}
suggesting a sequential importance sampling approach, where parameter particles are reweighted based on the incremental likelihood $p(y_t | y_{1:t-1}, \theta_{k}, \mathcal{M}_k)$.

For each of the $N_{\theta}$ $\theta$-particles, $\{\theta^{m}_{k}, \vartheta^{m}_{k,t} \}_{m=1}^{N_{\theta}}$ from $p(\theta_{k} | y_{1:t}, \mathcal{M}_k)$, we associate $N_x$ $x$-particles and perform the BPF. The SMC$^2$ algorithm reweights each $\theta$-particle according to an unbiased estimate of the incremental likelihood $p(y_t | y_{1:t-1}, \theta^{m}_{k}, \mathcal{M}_k)$, obtained from the BPF; see Equation \eqref{inc_bpf}. Here, $\{\vartheta^{m}_{k,t} \}_{m=1}^{N_{\theta}}$ denote the weights of the $\theta$-particles, which should not be confused with $\{w^{i}_{k,t} \}_{i=1}^{N_{x}}$, the weights of the state particles used in Algorithm~\ref{BMA-pf}. As time progresses, particle degeneracy may occur, whereby a few $\theta$-particles dominate the weight distribution. To mitigate this, a rejuvenation step is applied, consisting of a resampling phase followed by a move step using a particle marginal Metropolis-Hastings (PMMH) kernel \citep{Andrieu2010}. This move step aims to refresh the diversity of parameter particles by applying a few MCMC iterations to each resampled $\theta$-particle. These MCMC transitions are constructed to leave the target posterior distribution $p(\theta_k | y_{1:t}, \mathcal{M}_k)$ invariant.  The combination of the resampling and move steps is referred to as the ``rejuvenation'' step. In this study, new parameters are proposed from a multivarate Gaussian distribution $\theta^{*}_{k} \sim q(\theta^{*}_{k}|\theta^{m}_{k}):= \mathcal{N}(\widehat{\mu}_{k,t}, ~c\widehat{\Sigma}_{k,t})$, where $c$ is a scaling factor (set to $0.5$ in this work). The mean and covariance matrix are given by:
\begin{align}
 \widehat{\mu}_{k,t} = \dfrac{1}{\sum_{m=1}^{N_{\theta}} \vartheta^{m}_{k,t}} \sum_{m=1}^{N_{\theta}} \vartheta^{m}_{k,t} \theta^{m}_{k}, \text{ and }
 \widehat{\Sigma}_{k,t} = \dfrac{1}{\sum_{m=1}^{N_{\theta}} \vartheta^{m}_{k,t}} \sum_{m=1}^{N_{\theta}} \vartheta^{m}_{k,t} (\theta^{m}_{k}- \widehat{\mu}_{k,t})(\theta^{m}_{k} - \widehat{\mu}_{k,t})^\top,
\end{align}

It is worth noting that alternative proposals, such as a random walk proposal applied to the current set of particles, can also be used \citep{Chopin2020}.  The resampling in the rejuvenation step is similar to the one used in Algorithm \ref{BMA-pf} but is performed only when the effective sample size (ESS), which is an assessment of the degeneracy of the parameter particles, falls below a threshold $\tau_R$; in that case a degeneracy criterion is deemed to be met. The ESS for model $k$ is given by
\begin{align}
\text{ESS$_k$}=\left(\sum_{m=1}^{N_{\theta}}\vartheta^{m}_{k,t}\right)^2 \bigg/  \sum_{m=1}^{N_{\theta}}(\vartheta^{m}_{k,t})^2.
\end{align}
In practice, an ESS threshold of 50\% ($\tau_R=0.5$) is commonly used \citep{Chopin2013, Chopin2020}. 
Each rejuvenation step computes an unbiased estimate of the marginal likelihood for a proposed parameter $\theta^*_{k}$ by running Algorithm~\ref{BMA-pf} over observations from time 1 to $t$. While the computational load grows with the number of observations to process at each MCMC update, \cite{Chopin2013} show that the frequency of rejuvenation typically decreases over time, reducing the overall burden. The proposed parameter is then accepted or rejected according to the Metropolis-Hastings probability:
\begin{align}
\alpha_k = \min \left\{ 
1,\ 
\dfrac{ \widehat{p}_{_{N_x}}(y_{1:t} | \theta^*_{k}, \mathcal{M}_k)\ p(\theta^*_{k}) }
     { \widehat{p}_{_{N_x}}(y_{1:t} |  \theta^{m}_{k}, \mathcal{M}_k)\ p(\theta^{m}_{k}) }
\times
\dfrac{ q(\theta^{m}_{k} |  \theta^*_{k}) }
     { q(\theta^*_{k} |  \theta^{m}_{k}) }
\right\}.
\end{align}
Let's assume that $\{\theta^{m}_{k}, \vartheta^{m}_{k,t-1}\}_{m=1}^{N_\theta}$ is a particle approximation of the posterior $p(\theta_{k} | y_{1:t-1},\mathcal{M}_k)$. Using this approximation, we can incorporate parameter uncertainty into the model evidence estimation (see \cite{Chopin2013} for technical details). Specifically, the evidence of model $\mathcal{M}_k$ at time $t$ in Equation \eqref{marg_evidence} can be approximated as the product of weighted averages of the incremental likelihoods given the parameter particles:
\begin{align}\label{modelev}
p(y_{1:t} | \mathcal{M}_k) \approx \widehat{Z}_{k,t}
= \prod_{s=1}^{t} \left(
\dfrac{1}{\sum_{m=1}^{N_{\theta}} \vartheta^{m}_{k,s-1}}
\sum_{m=1}^{N_{\theta}} \vartheta^{m}_{k,s-1} \,
\widehat{p}_{_{N_x}}(y_s | y_{1:s-1}, \theta^{m}_{k}, \mathcal{M}_k)
\right),
\end{align}
where $\widehat{p}_{_{N_x}}(y_s | y_{1:s-1}, \theta^{m}_{k}, \mathcal{M}_k)$ is the estimated likelihood at time $s$ computed via the particle filter in \eqref{inc_bpf}.
 
For many infectious diseases, the relative adequacy of competing models may vary across different phases of an epidemic, such as transitions from early growth to peak incidence or from epidemic to controlled phases. These changes in model performance do not imply that the underlying parameters of the disease are changing; rather, different models may better capture the dynamics during specific phases. To account for this, we adopt a sliding window approach for computing model evidence. Importantly, this window does not impose temporal dynamics on the parameters themselves; it only evaluates each model’s posterior predictive performance over the most recent $t_w$ observations \citep[see also Section 5,][]{martino2017cooperative}. Under this approach, the model evidence for model $\mathcal{M}_k$ at time $t$ is approximated by
\begin{align}\label{modelev_tw}
\widehat{Z}^{(t_w)}_{k,t} = \prod_{s=t-t_w+1}^{t} \left(
\dfrac{1}{\sum_{m=1}^{N_{\theta}} \vartheta^{m}_{k,s-1}}
\sum_{m=1}^{N_{\theta}} \vartheta^{m}_{k,s-1} 
\widehat{p}_{_{N_x}}(y_s | y_{1:s-1}, \theta^{m}_{k}, \mathcal{M}_k)
\right).
\end{align}
Here, $\widehat{Z}^{(t_w)}_{k,t}$ represents the estimated model evidence based solely on observations $y_{t-t_w+1:t}$. Posterior model probabilities are then updated according to Equation~\eqref{true_pi}, yielding the approximation
\begin{align}\label{ck_tw}
\pi_{k,t} \approx \dfrac{\widehat{Z}^{(t_w)}_{k,t}P(\mathcal{M}_k)}{\sum_{j=1}^{K} \widehat{Z}^{(t_w)}_{j,t}P(\mathcal{M}_j)}, \quad k = 1,\ldots,K.
\end{align}
This sliding window approach evaluates model evidence using only the most recent $t_w$ observations, allowing model probabilities to adapt to changes in relative model performance over time. Models that better predict recent data receive higher posterior probabilities, while older observations have less influence. Without the sliding window, the impact of older data could overshadow the comparative performance of models in more recent periods. This approach is conceptually related to forgetting-factor formulations in dynamic model averaging \citep{raftery2010online}, in which contributions from earlier likelihood terms $p(y_{1:t-1} | \mathcal{M}_k)$ are down-weighted when updating model evidence. The SMC$^2$ algorithm for Bayesian model averaging with the sliding window is outlined in Algorithm~\ref{BMA-SMC2}.

Given that both the DTHP and SEIR models provide filtered estimates of the daily incidence and the reproduction number, the model-averaged estimates at each time step, denoted as $\widehat{\lambda}_t$ for the daily incidence and $\widehat{R}_t$ for the reproduction number, are calculated by accounting for both state and parameter uncertainty. To efficiently propagate parameter uncertainty without excessive computational cost, we select a subset of $N_c$ parameter particles $\{\theta^j_{k}\}_{j=1}^{N_c}$ from the full SMC$^2$ output according to their normalized weights. This subset provides a practical approximation of the posterior while keeping the computation tractable. For the model-averaged estimates, we then combine both state and parameter uncertainty as follows:
\begin{align}
\widehat{\lambda}_t &= \sum_{k=1}^2 \pi_{k,t} \widehat{\lambda}_{k,t}, \quad 
\mathrm{E}[\lambda_{k,t} | y_{1:t}] \approx  \widehat{\lambda}_{k,t} =\dfrac{1}{N_cN_x}  \sum_{j=1}^{N_c} \sum_{i=1}^{N_x} \lambda^{i}_{k,t,j} , \label{averge_inf} \\
\widehat{R}_t &= \sum_{k=1}^2 \pi_{k,t} \widehat{R}_{k,t}, \quad 
\mathrm{E}[R_{k,t} | y_{1:t}] \approx\widehat{R}_{k,t} =  \dfrac{1}{N_cN_x}  \sum_{j=1}^{N_c} \sum_{i=1}^{N_x} R^{i}_{k,t,j}. \label{averge_rt}
\end{align}
Here, $\{\lambda^{i}_{k,t,j}\}$ and $\{R^{i}_{k,t,j}\}$ represent the particle simulation of the incidence and the reproduction number, respectively, for model $\mathcal{M}_k$ based on data up to time $t$, given the $j$-th parameter particle $\theta^j_{k,t}$. This formulation explicitly accounts for both state uncertainty (via the particle filter) and parameter uncertainty (via the subset of $\theta$-particles), providing a robust and coherent approximation of $\lambda_t$ and $R_t$. Using $N_c = 100$ in our numerical experiments in Section~\ref{sec5} achieves a balance between computational efficiency and adequately capturing posterior variability, ensuring that the model-averaged estimates reflect the overall uncertainty without undue computational cost.

\begin{algorithm}[H]
\caption{Bayesian model averaging using SMC$^2$ with sliding window (BMA-SMC$^2$)}\label{BMA-SMC2}
\begin{flushleft}
Operations involving indexes $k$ and $m$ must be performed for $k = 1,\ldots,K$ and $m = 1, \dots, N_{\theta}$.

\textbf{Inputs:} Observations $y_{1:T}$, number of state particles $N_x$, number of parameter particles $N_\theta$, resample threshold $\tau_R$, number of PMMH moves $P$, sliding window size $t_w$, subset size for $\theta$-particles $N_c$.  

\textbf{Output:} Particle set $\{\theta^{m}_{k,}, x^{1:N_x}_{k,0:t,m}, \pi_{k,0:t}\}$.
\end{flushleft}
\hrulefill
\begin{algorithmic}[1]
\State Sample initial parameter particles $\theta^{m}_{k} \sim p(\theta_k)$ and set $\vartheta^m_{k,0} = 1/N_\theta$.
\For{$t = 1$ to $T$}
    \State Run Algorithm~\ref{BMA-pf} for each $\theta^{m}_{k}$ to obtain $\{x^{1:N_x}_{k,t,m}, w^{1:N_x}_{k,t,m}\}$ and incremental likelihood $\widehat{p}_{_{N_x}}(y_t | y_{1:t-1}, \theta^{m}_{k}, \mathcal{M}_k)$.
    \State Update parameter weights: $\vartheta^m_{k,t} = \vartheta^m_{k,t-1} \, \widehat{p}_{_{N_x}}(y_t | y_{1:t-1}, \theta^{m}_{k})$, then normalize.
    \If{ESS$_k < \tau_R N_\theta$}
        \State Resample parameter indices $a_{1:N_\theta} \sim \text{Resample}(\vartheta^{1:N_\theta}_{k,t})$.
        \State Set $\{\theta^{m}_{k}, \vartheta^m_{k,t}\} := \{\theta^{a_m}_{k,t}, 1/N_\theta\}$, $\{x^{1:N_x}_{k,t,m}, w^{1:N_x}_{k,t,m}\} := \{x^{1:N_x}_{k,t,a_m}, w^{1:N_x}_{k,t,a_m}\}$.
        \For{$p = 1$ to $P$} \Comment{PMMH rejuvenation step}
            \State Propose $\theta^*_k \sim q(\theta^*_k | \theta^{m}_{k})$.
            \State Run Algorithm~\ref{BMA-pf} with $\theta^*_k$ and $y_{1:t}$ to obtain $\{x^{1:N_x}_{k,t,*}, w^{1:N_x}_{k,t,*}\}$ and $\widehat{p}_{_{N_x}}(y_{1:t} | \theta^*_k, \mathcal{M}_k)$.
            \State Compute acceptance probability 
            \[
            \alpha_k = \min \Bigg\{1, \dfrac{\widehat{p}_{_{N_x}}(y_{1:t} | \theta^*_k) p(\theta^*_k)}{\widehat{p}_{_{N_x}}(y_{1:t} | \theta^{m}_{k}) p(\theta^{m}_{k})} \times
            \dfrac{q(\theta^{m}_{k} | \theta^*_k)}{q(\theta^*_k | \theta^{m}_{k})} \Bigg\}.
            \]
            \State Accept or reject $\theta^*_k$ with probability $\alpha_k$, updating $(\theta^{m}_{k}, x^{1:N_x}_{k,t,m}, w^{1:N_x}_{k,t,m})$.
        \EndFor
    \EndIf
    \State Compute sliding-window model evidence $\widehat{Z}^{(t_w)}_{k,t}$ using Eq.~\eqref{modelev_tw}  and estimate posterior model probability $\pi_{k,t}$ using Eq.~\eqref{ck_tw}.
    \EndFor
Select a subset of $N_c$ parameter particles $\{\theta^j_{k}\}_{j=1}^{N_c}$ according to normalized weights $\vartheta^j_{k,t}$ and Run Algorithm~\ref{BMA-pf} for each selected $\theta^j_{k}$ to obtain state particles $\{x^{1:N_x}_{k,t,j}, w^{1:N_x}_{k,t,j}\}$.
Compute model-averaged estimates incorporating both state and parameter uncertainty:
\[
\mathrm{E}[\Delta_t| y_{1:t}] \approx \widehat{\Delta}_t = \sum_{k=1}^{K} \pi_{k,t} \left(\dfrac{1}{N_cN_x}  \sum_{j=1}^{N_c}\sum_{i=1}^{N_x} \Delta^i_{k,t,j} \right),
\]
where $\Delta_t$ is one of $\lambda_t$ or $R_t$, i.e. a common quantity of interest across the models.
\end{algorithmic}
\end{algorithm}

A key advantage of sequential methods lies in their inherently parallel architecture. Most computations, such as particle propagation and weight updates, can be performed independently across particles. This structure enables efficient parallelization, allowing Algorithm~\ref{BMA-SMC2} to scale linearly with the number of state particles ($N_x$) and parameter particles ($N_\theta$). The algorithm is implemented in Python (version 3.10.12) and was executed on a desktop computer with an Intel Core i7-1300H processor running at 3.40 GHz. The PMMH updates of the $N_\theta$ parameter particles were parallelized across 10 cores.



\section{Results}\label{sec4}

\subsection{Experimental setup}

We assess the performance of the proposed BMA–SMC$^2$ framework in estimating daily incidence and the time-varying reproduction number using a controlled simulation study. The analysis has two aims: (i) to examine whether model averaging improves estimation and short-term forecasting relative to the individual DTHP and SEIR models; and (ii) to evaluate performance across epidemiological settings that vary in smoothness and predictability. Synthetic data are generated over a 100-day period with population size $N = 50000$. For SEIR-generated scenarios, the daily observed cases correspond to the number of individuals transitioning from $E$ to $I$, with mean incubation and infectious periods fixed at $1/\sigma = 2$ and $1/\gamma = 6$ days, respectively. Model filtered estimates use data up to $T=79$, after which forecasts are generated for an additional 21 days.

To reflect a broad spectrum of epidemic behaviours, three scenarios were constructed, differing in both smoothness and structural assumptions. Two scenarios are generated from SEIR dynamics, while a third is generated from the DTHP to explore performance when the data-generating process differs from the SEIR structure. In all SEIR scenarios, the initial state is $(S(0),E(0),I(0),R(0))=(N-10,0,10,0)$. For both models across all scenarios, we specify priors $\nu_k \sim \mathcal{TN}_{[0.05, 0.2]}(0.1, 0.01^2)$ and $\phi_k \sim \mathcal{U}([0, 0.2])$ for $k=1,2$, where $\mathcal{TN}_{[\text{inf, sup}]}$(mean, std$^2$) denotes a truncated normal distribution and $\mathcal{U}([a,b])$ denotes a uniform distribution.

\begin{enumerate}

\item \textbf{Scenario A} features a smoothly varying transmission rate that evolves as $\beta_t = 0.28 \exp\!\left(\cos\left(\frac{2\pi t}{96}\right) - \frac{t}{125}\right)$, yielding gradual undulations in transmission intensity with slowly decreasing amplitude. This provides a setting where changes occur continuously rather than abruptly, allowing assessment of model behaviour under stable and predictable dynamics. For the DTHP model, we set priors $\omega \sim \mathcal{TN}_{[0, 1]}(0.15, 0.05^2)$, $R_0 \sim \mathcal{U}([4, 4.5])$, and $\lambda_{H}(0) \sim \mathcal{U}(\{0, \dots, 5\})$. For the SEIR model, priors are $\sigma \sim \mathcal{TN}_{[0, 1]}(0.45, 0.1^2)$, $\gamma \sim \mathcal{TN}_{[0, 0.2]}(0.15, 0.05^2)$, $\beta_{0} \sim \mathcal{U}([0.7, 0.75])$, with initial conditions $(S(0), E(0), I(0), R(0)) = (N - E_0 - I_0, E_0, I_0, 0)$ where $E_0 \sim \mathcal{U}(\{0, \dots, 5\})$ and $I_0 \sim \mathcal{U}(\{0, \dots, 15\})$.

\item \textbf{Scenario B} is designed to test forecasting performance immediately prior to a rapid increase in the number of cases. The transmission rate $\beta_t$ follows a piecewise trajectory: $\beta_t = 0.35$ for $t \leq 40$, decreases linearly to $0.1$ at $t=41$, remains at $0.1$ until $t=80$, then increases linearly to $0.29$ by $t=93$ and stays constant thereafter. This pattern generates mild early transmission followed by a substantial resurgence. DTHP priors are $\omega \sim \mathcal{TN}_{[0, 1]}(0.1, 0.05^2)$, $R_0 \sim \mathcal{U}([1.8, 2.1])$, and $\lambda_{H}(0) \sim \mathcal{U}(\{0, \dots, 5\})$. SEIR priors are $\sigma \sim \mathcal{TN}_{[0, 1]}(0.45, 0.1^2)$, $\gamma \sim \mathcal{TN}_{[0, 1]}(0.15, 0.05^2)$, and $\beta_{0} \sim \mathcal{U}([0.33, 0.37])$, with the same initial condition specification as Scenario A.

\item \textbf{Scenario C} considers observed cases generated from the DTHP model with a geometric triggering kernel and no background rate ($\mu = 0$). The reproduction number evolves as: $R_t = 1.5$ for $t \leq 30$, decreases linearly to $0.85$ for $30 < t \leq 55$, and remains at $0.80$ for $t > 55$, combined with decay parameter $\omega = 0.2$ and $10$ initial cases. This generates a controlled decline in transmissibility where the data-generating process differs fundamentally from the SEIR structure. DTHP priors are $\omega \sim \mathcal{TN}_{[0, 1]}(0.15, 0.1^2)$, $R_0 \sim \mathcal{U}([1.35, 1.42])$, and $\lambda_{H}(0) \sim \mathcal{U}(\{0, \dots, 5\})$. SEIR priors are $\sigma \sim \mathcal{U}([0, 1])$, $\gamma \sim \mathcal{U}([0, 1])$, and $\beta_{0} \sim \mathcal{U}([0.27, 0.37])$, with the same initial condition specification as previous scenarios.

\end{enumerate}

\subsection{Evaluation Metrics}

To evaluate model performance for both in-sample estimation and out-of-sample forecasting, we employ three complementary metrics: Root Mean Square Error (RMSE), coverage of predictive credible intervals (\text{Coverage}$_{\alpha}$), and the Continuous Ranked Probability Score (CRPS). 
Let $\mathcal{T}$ denote the set of time points over which a metric is computed. Depending on the context, this may correspond to the in-sample period ($t=1,\dots,T$) or the forecast period ($t=T+1,\dots,T+H$), where $H$ is the forecasting horizon.

RMSE measures the average discrepancy between the model-predicted mean and the true value. Denoting the predictive mean at time $t$ by $\widehat{\Delta}_t \approx \mathrm{E}[\Delta_t | y_{1:t}]$ for in-sample evaluation  ($t\leq T$) or $\widehat{\Delta}_{T+h} \approx \mathrm{E}[\Delta_{T+h} | y_{1:T}]$ for an $h$-step-ahead forecast ($1\leq h \leq H$), RMSE is defined as: 
\begin{align*} 
\text{RMSE} = \sqrt{\dfrac{1}{|\mathcal{T}|} \sum_{t \in \mathcal{T}} (z_t - \widehat{\Delta}_t)^2}, 
\end{align*}
where $z_t$ is the observed number of cases (or reproduction number, which we will know for simulated data) at time $t$.

Coverage quantifies how well model uncertainty reflects the true variability of the data. For each $t \in \mathcal{T}$, we construct a $(1-\alpha)$ credible interval for $\Delta_t$ from the filtering distribution (in-sample) or the forecast distribution (out-of-sample).
The empirical coverage is  
\begin{align*}
    \text{Coverage}_{\alpha} = \dfrac{1}{|\mathcal{T}|} \sum_{t \in \mathcal{T}} 
    \mathbb{I}\big(z_t \in \mathrm{CI}_{\alpha,t}\big),
\end{align*}
where $\mathrm{CI}_{\alpha,t}$ denotes the corresponding credible interval.In our results, we use $1-\alpha = 0.95$.

CRPS evaluates the entire predictive distribution rather than only its mean or an interval. Let $F_t(u)$, $t \in \mathcal{T}$ denote the predictive cumulative distribution function at time $t$, given by $F_t(u)=p(\Delta_t \leq u | y_{1:t})$ in-sample ($t\leq T$)  or $F_{T+h}(u)=p(\Delta_{T+h} \leq u | y_{1:T})$ for forecasts ($1\leq h \leq H$) .
Then
\begin{align*}
    \text{CRPS} = 
    \dfrac{1}{|\mathcal{T}|} \sum_{t \in \mathcal{T}}
    \int_{-\infty}^{\infty} \big(F_t(u) - \mathbb{I}(z_t \leq u)\big)^2 \, du.
\end{align*}

In practice, the predictive distribution at time $t$ is represented by a joint collection of particles $\{\Delta_t^{(p)}\}_{p=1}^{N_x N_c}:=\{\Delta_{t,j}^{i}, ~i=1,\dots, N_x \text{ and } j=1,\dots, N_c \}$ obtained by sampling $N_x$ state particles for each of the $N_c$ parameter particles retained.
Using this particle approximation, CRPS is computed as
\begin{align*}
  \text{CRPS} \approx \dfrac{1}{|\mathcal{T}|} \sum_{t \in \mathcal{T}} 
  \left(
  \dfrac{1}{N_x N_c} \sum_{p=1}^{N_x N_c} |z_t - \Delta_t^{(p)}|
  - \dfrac{1}{2 (N_x N_c)^2} 
    \sum_{p=1}^{N_x N_c} \sum_{q=1}^{N_x N_c} |\Delta_t^{(p)} - \Delta_t^{(q)}|
  \right),
\end{align*}
which captures both the bias and the dispersion of the filtering distribution \citep{jordan2019evaluating}.


For the purposes of this study, we set $\mu = 0$ in Equation \eqref{dthp} of the DTHP, so that all new cases arise solely from transmission by previously infected individuals within the population. This ensures consistency with the SEIR model, which also does not explicitly account for imported or spontaneous cases.  For each model, Algorithm \ref{BMA-SMC2} was executed with $N_{\theta} = 400$ parameter particles and $N_x = 200$ state particles, with $P=5$ successive PMMH moves. We employed a sliding window of one time step ($t_w = 1$) for computing the model evidence (see Equation \eqref{modelev_tw}), meaning that the models are weighted based on the one-step-ahead predictive likelihood \citep{parrish2012toward, martino2017cooperative}. This choice was motivated by preliminary simulation results (not shown), which indicated that using a one-step window provides slightly better predictions for model averaging compared to using the full model evidence as in Equation \eqref{modelev}.

\begin{figure}[H]
    \centering
    \includegraphics[width=1\linewidth]{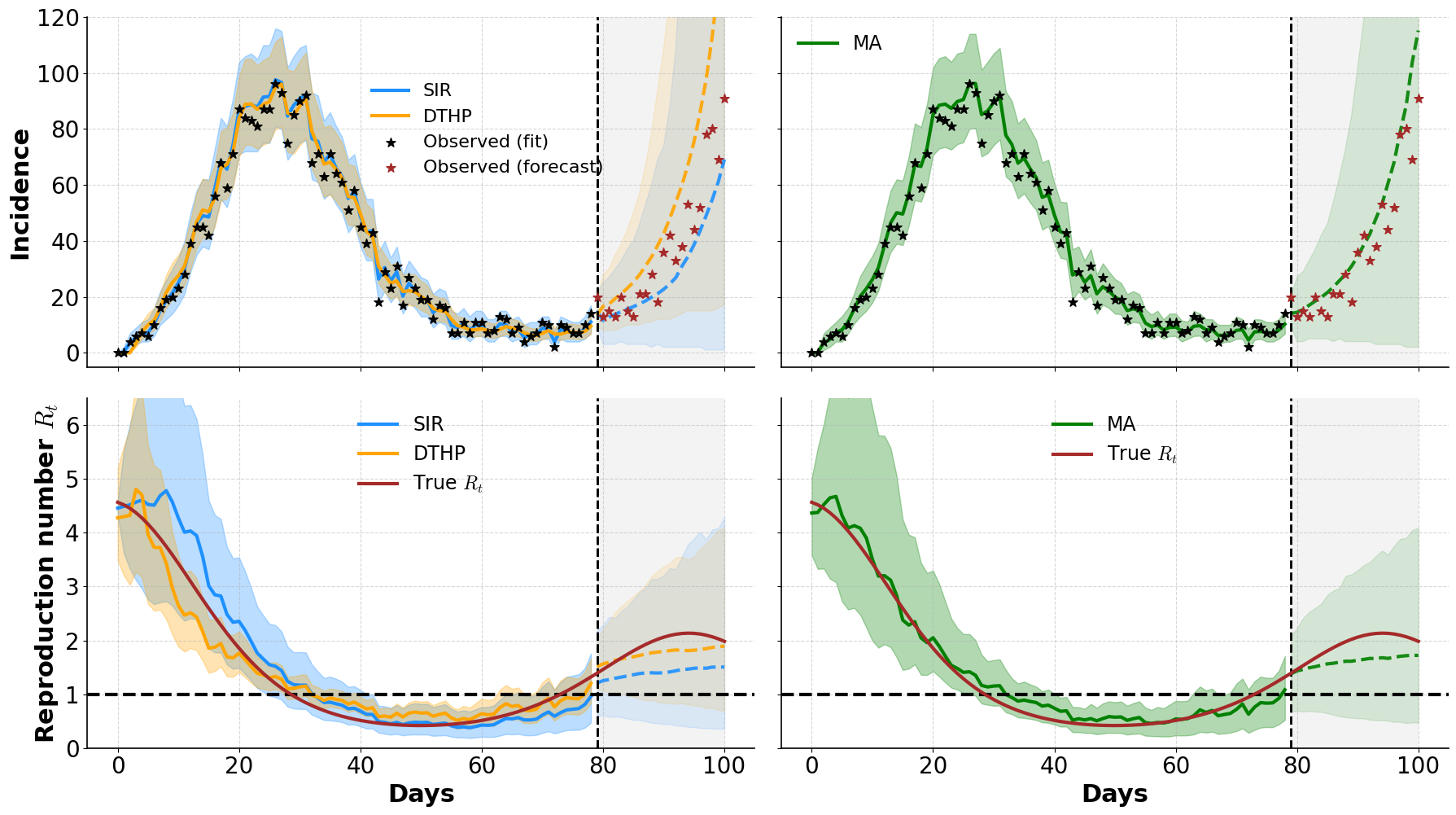}
\caption{\footnotesize Simulated daily incidence and time-varying reproduction number for Scenario A. The black/brown stars correspond to the observed incidence data, and the true underlying $R_t$ is shown as a solid brown line. The solid lines represent the posterior mean estimates of the incidence and $R_t$ obtained from the DTHP (orange), SEIR (blue), and MA (green) approaches, with shaded areas indicating the associated 95\% credible intervals. The vertical black dashed line marks the start of the forecasting period.}
    \label{Fig1}
\end{figure}

\begin{figure}[H]
    \centering
    \includegraphics[width=1\linewidth]{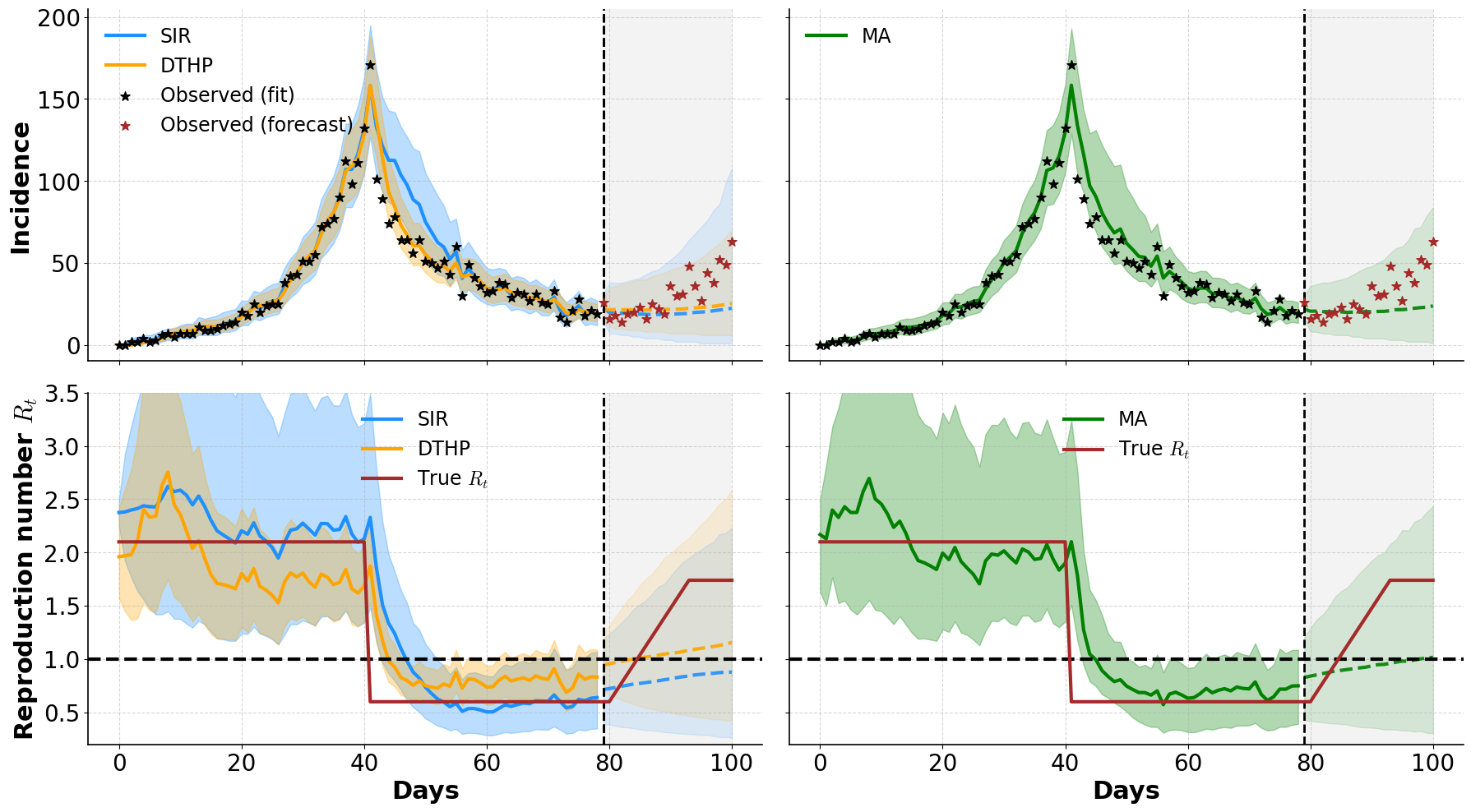}
\caption{\footnotesize Simulated daily incidence and time-varying reproduction number for Scenario B. The black/brown stars correspond to the observed incidence data, and the true underlying $R_t$ is shown as a solid brown line. The solid lines represent the posterior mean estimates of the incidence and $R_t$ obtained from the DTHP (orange), SEIR (blue), and MA (green) approaches, with shaded areas indicating the associated 95\% credible intervals. The vertical black dashed line marks the start of the forecasting period.}
    \label{Fig2}
\end{figure}

\begin{figure}[H]
    \centering
    \includegraphics[width=1\linewidth]{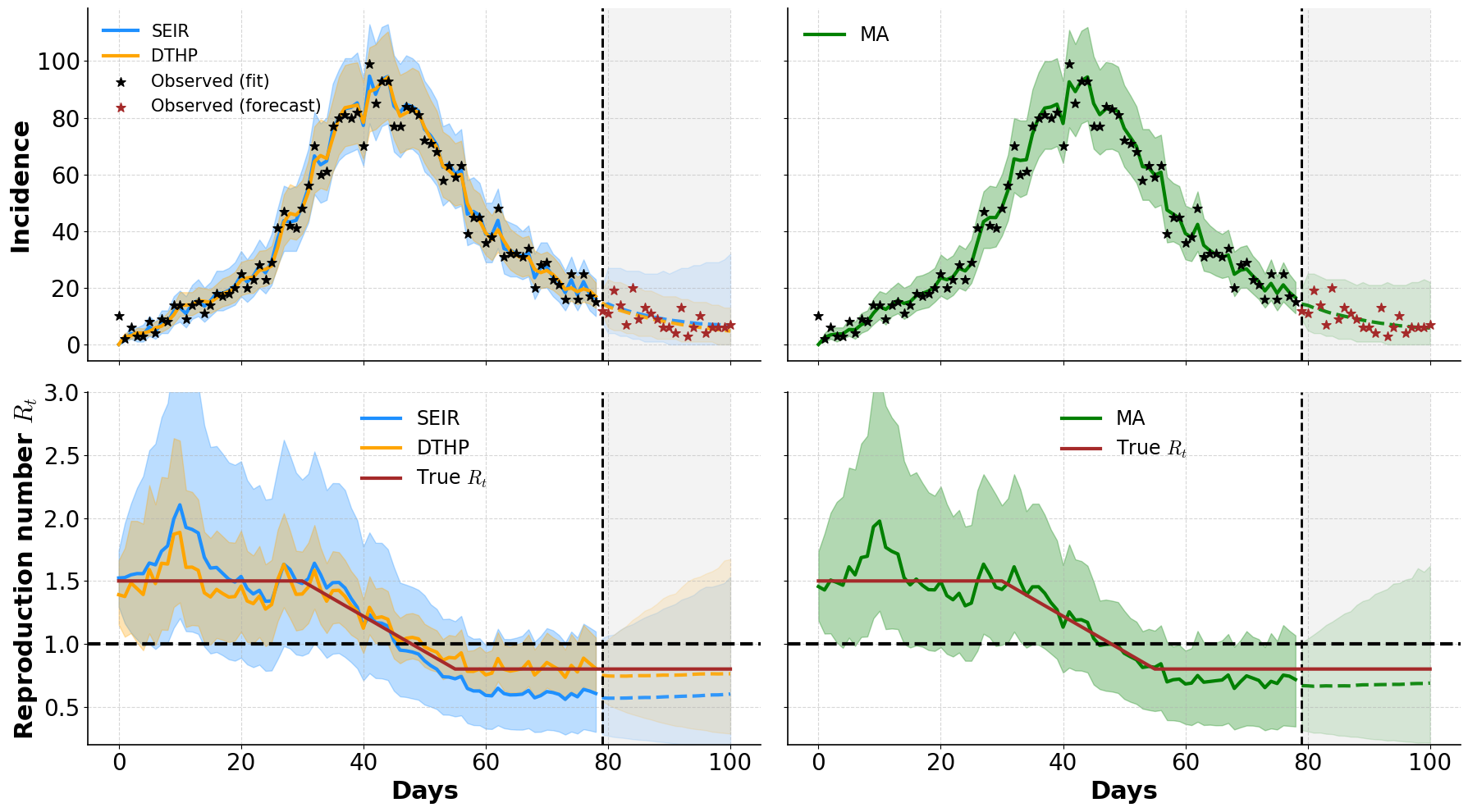}
\caption{\footnotesize Simulated daily incidence and time-varying reproduction number for Scenario C. The black/brown stars correspond to the observed incidence data, and the true underlying $R_t$ is shown as a solid brown line. The solid lines represent the posterior mean estimates of the incidence and $R_t$ obtained from the DTHP (orange), SEIR (blue), and MA (green) approaches, with shaded areas indicating the associated 95\% credible intervals. The vertical black dashed line marks the start of the forecasting period.}

    \label{Fig3}
\end{figure}

Figures~\ref{Fig1}–\ref{Fig3} compare the three approaches: DTHP, SEIR, and Model Averaging (MA) across Scenarios~A–C in terms of their ability to estimate daily incidence and track the time-varying reproduction number $R_t$. Shaded regions denote 95\% credible intervals, representing posterior uncertainty. Quantitative performance is summarised using within-sample and out-of-sample metrics in Tables~\ref{Tab1} and~\ref{Tab2}, respectively.

In Scenario A (Figure~\ref{Fig1}), which represents smoothly varying transmission dynamics, all three models provide accurate fits to the observed incidence during the training period. As shown in Table~\ref{Tab1}, SEIR achieves the lowest within-sample RMSE and CRPS for incidence, alongside full empirical coverage. MA performs comparably in terms of CRPS and also attains full coverage. For $R_t$, MA yields the lowest RMSE and CRPS, indicating improved accuracy and sharper inference relative to the individual models, though DTHP exhibits lower coverage. In the forecasting period (Table~\ref{Tab2}), MA attains the lowest RMSE for incidence, though DTHP achieves lower CRPS, and both models maintain well-calibrated predictive intervals with full coverage alongside SEIR. For $R_t$, DTHP provides the most accurate forecasts in terms of both RMSE and CRPS, with all models achieving full coverage. The full in-sample and forecast computation, including the sequential model-averaging procedure, required approximately 28 minutes of CPU time for Scenario A.

Scenario B (Figure~\ref{Fig2}) examines forecasting performance immediately prior to a sharp increase in transmission, exploring how models perform under an abrupt epidemic turning point. During the fitting period, all three models track observed incidence reasonably well, with DTHP achieving the lowest RMSE, followed by MA and SEIR (Table~\ref{Tab1}). SEIR struggles to capture the rapid decline after day 41, resulting in larger errors. MA provides the best overall balance between accuracy and uncertainty, as reflected in its lowest CRPS for both incidence and $R_t$ during the fitting period. In the forecasting window (Table~\ref{Tab2}), none of the models anticipate the imminent surge as would be expected: Without new data indicating the upcoming change in transmission, no model, whether mechanistic or statistical, can reliably forecast such turning points. DTHP achieves slightly lower RMSE and CRPS for incidence compared to MA and SEIR, but forecast errors remain substantial across all approaches. For $R_t$, DTHP shows lower point estimate errors but slightly reduced coverage, while SEIR and MA maintain full coverage through wider, more conservative intervals. These results highlight a fundamental limitation of epidemic forecasts since models cannot anticipate turning points without early signals in the observed data. The complete computation for Scenario B, including forecasts, took approximately 1 hour 08 minutes of CPU time, reflecting the added complexity of abrupt transmission changes as parameter particles quickly degenerate.

Scenario C (Figure~\ref{Fig3}) considers a setting in which incidence is generated from a DTHP. During the training period, the DTHP model provides the most accurate estimates of $R_t$ in terms of both RMSE and CRPS, substantially outperforming SEIR, while MA shows intermediate performance (Table~\ref{Tab1}). Correspondingly, the DTHP model receives the highest posterior weight, as shown in Figure~\ref{FigB4} in Appendix~\ref{appB}. For incidence, SEIR achieves the lowest RMSE and CRPS despite the differing generative dynamics, with MA performing similarly. All models achieve full or near-full coverage for both variables. During the forecasting period (Table~\ref{Tab2}), MA yields the lowest RMSE for incidence, while the DTHP model achieves slightly lower CRPS. For $R_t$, the DTHP model again provides the most accurate forecasts, with RMSE even improving relative to within-sample performance. It should be noted that this improvement reflects the fact that $R_t$ remains relatively stable during the forecasting window, in contrast to the more variable dynamics present during the training period. All models maintain full predictive coverage in this scenario. The total CPU time for Scenario C, including both in-sample and forecast computations, was approximately 37 minutes.

\begin{table}[H]
\centering
\caption{Summary of within-sample model performance across Scenarios A, B, and C.}
\label{Tab1}
\begin{tabular}{cl*{10}{r}}
\toprule
\multirow{2}{*}{\textbf{Variable}} & \multirow{2}{*}{\textbf{Metric}} 
  & \multicolumn{3}{c}{\textbf{Scenario A}} 
  & \multicolumn{3}{c}{\textbf{Scenario B}} 
  & \multicolumn{3}{c}{\textbf{Scenario C}} \\
\cmidrule(lr){3-5} \cmidrule(lr){6-8} \cmidrule(lr){9-11}
 & & DTHP & SEIR & MA & DTHP & SEIR & MA & DTHP & SEIR & MA \\
\midrule
\multirow{3}{*}{Incidence} 
  & RMSE     
  & 3.941 & \textbf{3.048} & 3.325 
  & \textbf{7.073} & 11.481 & 7.552 
  & 3.896 & \textbf{2.737} & 3.220 \\
  & Coverage$_{\alpha}$ 
  & 0.861 & \textbf{0.987} & \textbf{0.987} 
  & 0.861 & 0.924 & \textbf{0.962} 
  & 0.911 & \textbf{0.987} & \textbf{0.987} \\
  & CRPS     
  & 2.218 & \textbf{1.614} & 1.781 
  & 2.892 & 4.345 & \textbf{2.868} 
  & 2.216 & \textbf{1.705} & 1.887 \\
\midrule
\multirow{3}{*}{$R_t$} 
  & RMSE     
  & 0.286 & 0.373 & \textbf{0.179} 
  & 0.341 & 0.375 & \textbf{0.314} 
  & \textbf{0.103} & 0.178 & 0.120 \\
  & Coverage$_{\alpha}$ 
  & 0.658 & \textbf{1.000} & \textbf{1.000} 
  & 0.544 & \textbf{0.911} & \textbf{0.911} 
  & \textbf{1.000} & \textbf{1.000} & \textbf{1.000} \\
  & CRPS     
  & 0.168 & 0.169 & \textbf{0.130} 
  & 0.213 & 0.176 & \textbf{0.168} 
  & \textbf{0.056} & 0.109 & 0.069 \\
\bottomrule
\end{tabular}
\end{table}

\begin{table}[H]
\centering
\caption{Summary of out-of-sample model performance across Scenarios A, B, and C.}
\label{Tab2}
\begin{tabular}{{cl}*{10}{r}}
\toprule
\multirow{2}{*}{\textbf{Variable}} & \multirow{2}{*}{\textbf{Metric}} 
  & \multicolumn{3}{c}{\textbf{Scenario A}} 
  & \multicolumn{3}{c}{\textbf{Scenario B}} 
  & \multicolumn{3}{c}{\textbf{Scenario C}} \\
\cmidrule(lr){3-5} \cmidrule(lr){6-8} \cmidrule(lr){9-11}
 & & DTHP & SEIR & MA & DTHP & SEIR & MA & DTHP & SEIR & MA \\
\midrule
\multirow{3}{*}{Incidence} 
  & RMSE     
  & 24.716 & 12.045 & \textbf{11.618} 
  & \textbf{14.964} & 16.477 & 15.668 
  & 3.528 & 3.461 & \textbf{3.459} \\
  & Coverage$_{\alpha}$ 
  & 0.955 & \textbf{1.000} & \textbf{1.000} 
  & \textbf{1.000} & \textbf{1.000} & \textbf{1.000} 
  & \textbf{1.000} & \textbf{1.000} & \textbf{1.000} \\
  & CRPS     
  & 8.322 & 12.187 & \textbf{8.228} 
  & \textbf{8.370} & 10.202 & 9.117 
  & \textbf{1.974} & 2.096 & 2.000 \\
\midrule
\multirow{3}{*}{$R_t$} 
  & RMSE     
  & \textbf{0.207} & 0.532 & 0.336 
  & \textbf{0.456} & 0.649 & 0.545 
  & \textbf{0.048} & 0.220 & 0.127 \\
  & Coverage$_{\alpha}$ 
  & \textbf{1.000} & \textbf{1.000} & \textbf{1.000} 
  & 0.909 & \textbf{1.000} & \textbf{1.000} 
  & \textbf{1.000} & \textbf{1.000} & \textbf{1.000} \\
  & CRPS     
  & \textbf{0.189} & 0.377 & 0.250 
  & \textbf{0.291} & 0.430 & 0.344 
  & \textbf{0.067} & 0.162 & 0.100 \\
\bottomrule
\end{tabular}
\end{table}

While no single model dominates across all settings or performance criteria, the results demonstrate that MA provides consistently reliable performance across a range of epidemiological scenarios. Individual models occasionally outperform MA when their assumptions align closely with the data-generating process, but MA remains competitive across all scenarios and metrics, offering a balanced compromise between accuracy, calibration, and adaptability. This is particularly evident in scenarios with non-stationary transmission dynamics, where MA achieves intermediate errors and maintains coverage comparable to the best-performing model. The near-full predictive coverage observed during the forecast period (Table~\ref{Tab2}) reflects the wider predictive intervals that result from the absence of new observations to inform state and parameter update; models adopt conservatively wide intervals to maintain calibrated uncertainty as the forecast horizon extends. The framework also allows dynamic tracking of model parameter distributions over time ($t = 1, \dots, T$), supporting real-time assessment of parameter uncertainty, as illustrated in Figures \ref{FigB1}--\ref{FigB3} in Appendix \ref{appB}.

\subsection{Analysis of real data: Irish Influenza and COVID-19 epidemics}
We applied our methodology to real-world data from Ireland, using weekly influenza cases and daily COVID-19 case counts. The data were obtained from the Health Protection Surveillance Centre (HPSC). For both applications, we employed $N_{\theta} = 500$ parameter particles and $N_x = 1000$ state particles, with $P=5$ successive PMMH moves. Prior distributions for model parameters and initial conditions are detailed in Appendix \ref{appC}.

For influenza, we analyzed weekly cases from May 29, 2022, to June 16, 2024, after most COVID-19 restrictions had ended, allowing for a clear assessment of post-pandemic influenza dynamics. Inference was performed using the methods described in Section \ref{sec3}, with an SEIRS model as one baseline (see Appendix \ref{appC}). Historical data from the first 94 weeks were used for sequential training, followed by 14-week-ahead forecasts.

\begin{figure}[H]
    \centering
   \includegraphics[width=1\linewidth]{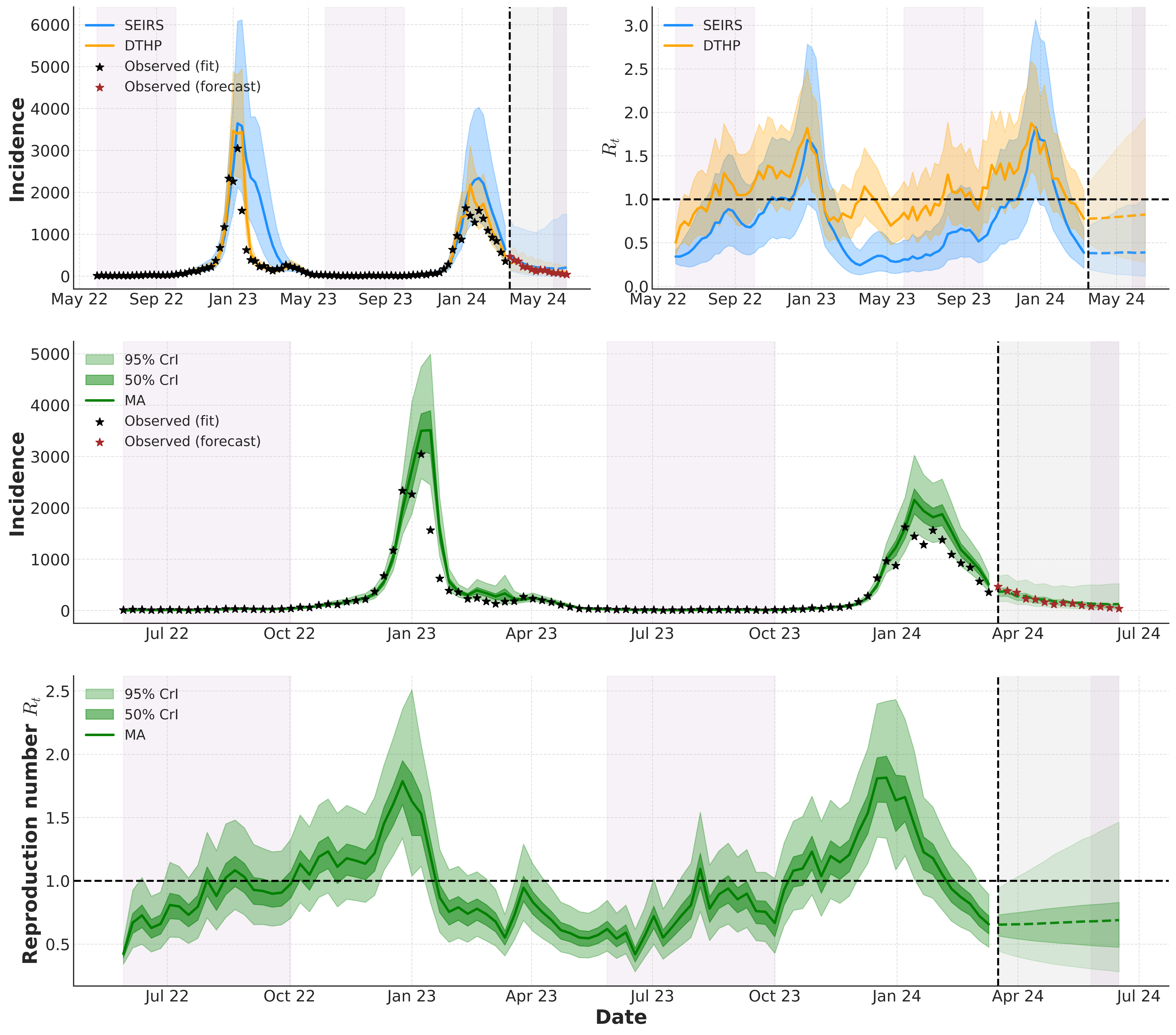}
      \caption{\footnotesize weekly incidence data from the Influenza pandemic and corresponding filtering estimate of the time-varying reproduction number for the DTHP and SEIRS models. The black/brown starts correspond to the observed data for incidence. The solid lines represent the posterior mean estimates of the incidence and $R_t$  obtained from the DTHP (orange), and SEIRS (blue) approaches, with shaded areas indicating the associated 95\% credible intervals. The vertical black dashed line marks the start of the forecasting period, and the light gray area highlights the forecast window. Summer periods are indicated by a light pink area. }
    \label{Fig4}
\end{figure}

Figure \ref{Fig4} presents model-filtered estimates of influenza incidence and the corresponding $R_t$ estimates for the SEIRS, DTHP, and MA approaches. The SEIRS model fails to fully capture the sharp decline in cases during the first wave and slightly overestimates the second wave peak. In contrast, the DTHP model closely tracks changes in observed incidence but exhibits lower coverage. Both models yield similar $R_t$ trends, with modest differences during periods of low incidence between peaks. The model posterior probabilities (Figure \ref{FigC3}) reveal that MA dynamically adapts between SEIRS and DTHP throughout the epidemic, favouring DTHP during periods of rapid transmission changes. As shown in Table \ref{Tab3}, MA achieves the lowest RMSE during the in-sample period, while DTHP attains the lowest CRPS. SEIRS maintains the highest coverage but at the cost of substantially larger errors. During the forecasting period from March 10 to June 16, 2024, MA outperforms both baseline models, achieving the lowest RMSE and CRPS while maintaining full coverage alongside DTHP and SEIRS.  The MA approach effectively balances the ability to capture abrupt changes (characteristic of DTHP) with the broader epidemic structure inferred by SEIRS, resulting in visually smoother estimates that follow the observed data closely. The lower out-of-sample RMSE relative to within-sample performance reflects the shorter forecast horizon and the positioning of the forecast window during a stable, declining phase of transmission. The $R_t$ estimates capture seasonal variations, with transmission peaking in winter: the 2022--2023 season, dominated by Influenza A, peaked at 1.78 (95\% CI: 1.34--2.35), while 2023--2024 peaked at 1.81 (95\% CI: 1.33--2.41) before declining by June 30, 2024. The full analysis, including both in-sample fitting and 14-week-ahead forecasts, required approximately 2 hours 20 minutes of CPU time.

\begin{figure}[H]
    \centering
    \includegraphics[width=1\linewidth]{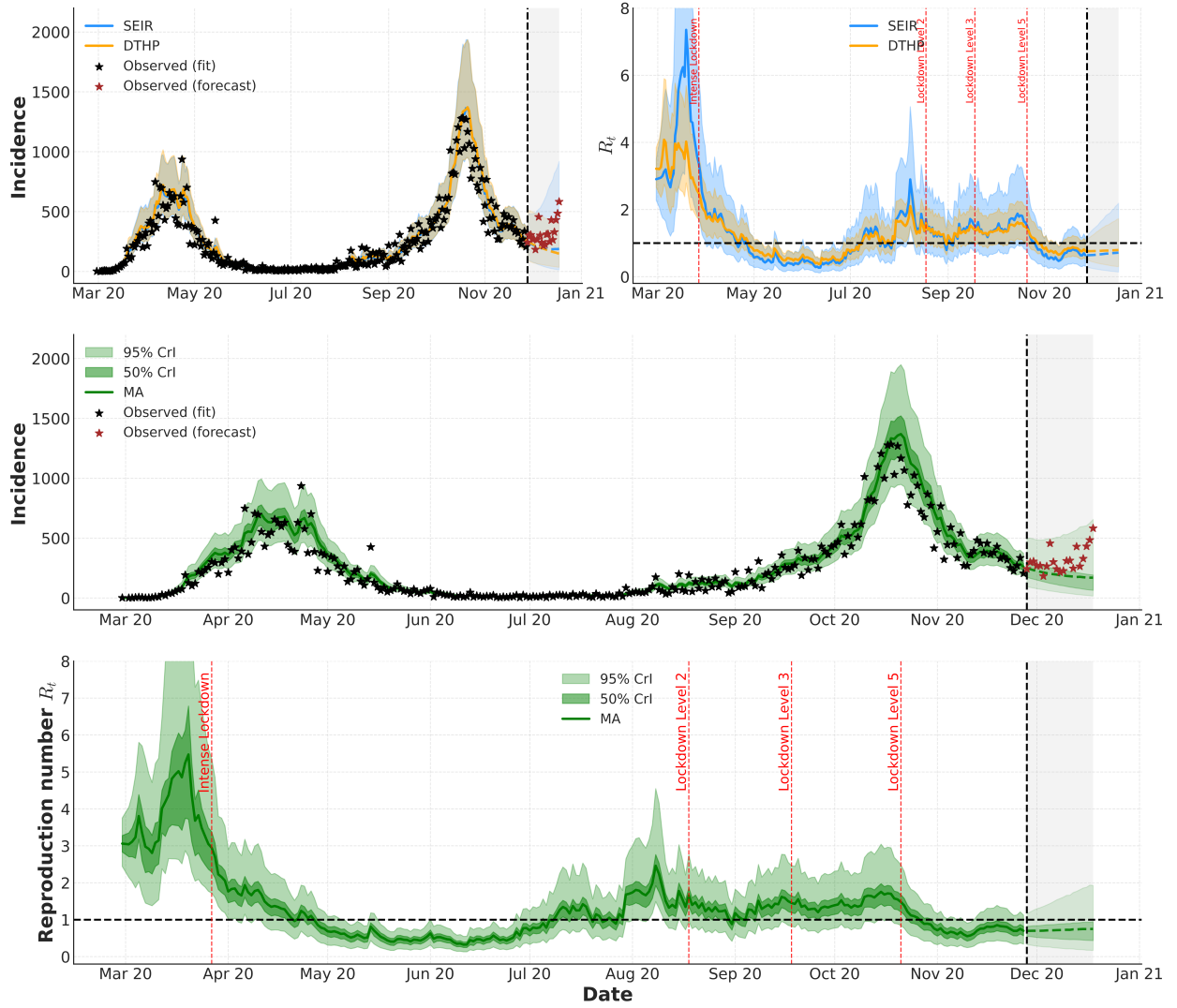}
\caption{\footnotesize Daily case incidence from the COVID-19 and corresponding filtering estimate of the time-varying reproduction number. The black/brown starts correspond to the observed data for incidence.  The solid lines represent the posterior mean estimates of the incidence and $R_t$ obtained from the DTHP (orange), SEIR (blue), and MA (green) approaches, with shaded areas indicating the associated 95\% credible intervals. Vertical red dashed lines mark the start dates of major mitigation measures, the vertical black dashed line indicates the beginning of the forecasting period, and the light gray area denotes the forecast window.}
\label{Fig5}
\end{figure}

For COVID-19, we focused on the initial waves from February 29 to December 18, 2020, a period marked by high transmission and changing interventions. Figure~\ref{Fig5} shows that all models closely fit the observed incidence, capturing the two epidemic waves. Visually, the DTHP model follows changes in incidence more closely than SEIR, though coverage is lower, while SEIR slightly overestimates early transmission and produces broader predictive intervals. Table~\ref{Tab3} confirms that during the in-sample period, SEIR achieves the lowest error metrics with the highest coverage, while DTHP demonstrates reduced coverage. For the 3-week forecasts from November 28, SEIR attains the lowest RMSE with full coverage, whilst MA achieves the lowest CRPS, also with full coverage. DTHP shows moderate performance with good but imperfect coverage. We further notice that all models still struggle to fully capture changes in the transmission process occurring during the forecast period, indicating that sudden shifts in transmission dynamics remain challenging to anticipate. The full inference and forecast procedure for COVID-19 required approximately 8 hours of CPU time. Both SEIR and DTHP models produce similar $R_t$ trends. The estimates reflect key epidemic dynamics, including a sharp decline following the first lockdown (March 27, 2020), which brought $R_t$ below one and ended the first wave. As restrictions eased in June, $R_t$ gradually increased and remained above one until mid-October, marking the second wave’s onset. Lockdowns reintroduced in late October brought $R_t$ below one again by November. Our $R_t$ estimates align with previous findings \citep{jaouimaa2021age, cazelles2021mechanistic}.

\begin{table}[H]
\centering
\caption{Summary of model performance on Influenza and COVID-19 data.}
\label{Tab3}
\begin{tabular}{{ll}*{6}{r}}
\toprule
\textbf{Disease} & \textbf{Metric} 
  & \textbf{DTHP} & \textbf{SEIR(S)} & \textbf{MA} 
  & \textbf{DTHP} & \textbf{SEIR(S)} & \textbf{MA} \\
 \cmidrule(lr){3-5} \cmidrule(lr){6-8}
& & \multicolumn{3}{c}{In-sample} & \multicolumn{3}{c}{Out-of-sample} \\
\midrule

\multirow{3}{*}{Influenza} 
  & RMSE 
    & 280.363 & 536.930 & \textbf{275.273} 
    & 63.791 & 88.958 & \textbf{58.258} \\
  & Coverage$_{\alpha}$ 
    & 0.798 & \textbf{0.883} & 0.830 
    & \textbf{1.00} & \textbf{1.00} & \textbf{1.00} \\
  & CRPS 
    & \textbf{69.570} & 160.448 & 73.504 
    & 33.286 & 41.330 & \textbf{27.865} \\
\midrule

\multirow{3}{*}{COVID-19} 
  & RMSE 
    & 91.950 & \textbf{82.708} & 87.465 
    & 172.408 & \textbf{160.116} & 166.281 \\
  & Coverage$_{\alpha}$ 
    & 0.651 & \textbf{0.835} & 0.801 
    & 0.810 & \textbf{1.00} & \textbf{1.00} \\
  & CRPS 
    & 40.050 & \textbf{34.526} & 37.087 
    & 106.503 & 106.954 & \textbf{106.393} \\
\bottomrule
\end{tabular}

\vspace{1ex}
\begin{minipage}{0.9\linewidth}
\footnotesize \textit{Note:} For Influenza, the SEIRS model was used in place of the SEIR model.
\end{minipage}
\end{table}


\section{Discussion}\label{sec5}
Dynamic models are essential tools for monitoring disease progression and anticipating new epidemic waves. In this study, we introduced a novel approach to enhance epidemic modeling by combining the strengths of the DTHP and SEIR(S) models through a model-averaging (MA) technique. Our method uses the SMC$^2$ framework to dynamically integrate predictions from multiple models, yielding a single, robust estimate of shared latent variables while simultaneously estimating static parameters.

Our results demonstrate that both the DTHP and SEIR(S) models are individually able to capture epidemic trends and estimate the time-varying reproduction number, $R_t$, under diverse conditions. The DTHP model, which captures self-exciting behavior, performed well during phases marked by abrupt transmission changes. However, it produced less accurate estimates of $R_t$, likely due to the limitations of its simplified triggering kernel. In contrast, the SEIR(S) model more accurately represented the overall epidemic trajectory and the evolution of $R_t$, but struggled to capture sudden changes in incidence. Additionally, it produced wider credible intervals, possibly reflecting uncertainty in the infectious period, which directly influences $R_t$ estimates. These observations underscore the limitations of relying on a single model, particularly in complex and rapidly evolving epidemic contexts.

The proposed model-averaging framework balances the complementary strengths of the DTHP and SEIR(S) models, producing well-calibrated forecasts with appropriate uncertainty coverage. It performs robustly in both in-sample and out-of-sample settings, consistently yielding the lowest or near-lowest prediction errors for incidence and $R_t$ across various simulation scenarios. It also reduces prediction uncertainty, producing more balanced credible intervals while maintaining high predictive accuracy. It is important to note that the out-of-sample evaluation focused primarily on periods of relatively stable transmission. In periods where the transmission process changes rapidly within the forecast horizon, all models, including MA, struggled to fully anticipate these shifts. Exploratory forecasts immediately prior to sharp increases in incidence showed higher errors across all approaches. This limitation reflects the inherent difficulty of forecasting abrupt turning points without early warning signals in the data and represents an important challenge for real-time forecasting applications, where early detection of epidemic turning points remains an open problem. Consequently, while MA generally improves stability, calibration, and uncertainty quantification, its out-of-sample predictions remain sensitive to rapid or unexpected shifts in epidemic dynamics. We further demonstrated the practical utility of this framework using empirical data from influenza and COVID-19 epidemics in Ireland. During influenza waves with abrupt changes in transmission, the MA approach successfully combined the benefits of DTHP and SEIR(S), delivering accurate forecasts and reliable estimates. For COVID-19, although improvements were more modest, model averaging yielded more stable $R_t$ estimates during periods of changing transmission and handled uncertainty more effectively. These findings suggest that our sequential model-averaging strategy is a promising approach for real-time epidemic surveillance and forecasting, offering a valuable balance between predictive performance and uncertainty quantification, even when the forecast horizon includes non-stationary transmission dynamics. The sequential ensemble framework further enhances flexibility by continuously updating posterior model probabilities, allowing adaptation to shifting epidemic dynamics (see Figures \ref{FigB4} in Appendix \ref{appB} and \ref{FigC3} in Appendix \ref{appC}). This adaptability is particularly valuable in situations where distinct phases of the epidemic may favor different modeling approaches. Note that this comparison focuses on epidemiological modeling using case incidence data and should not be interpreted as a general assessment of the overall performance of individual models.

While model averaging enhances predictive accuracy and robustness, it also introduces challenges. One is interpretability: the combined model's behavior is often harder to explain than that of its individual components, complicating communication with public health stakeholders. Still, interpretability is partly retained through outputs such as SEIR-derived estimates of latency and infectious periods, or DTHP-triggering kernels that quantify how past infection events influence expected case counts \citep{browning2021simple}. These elements provide valuable mechanistic insights into the epidemic process (see Figures \ref{FigC1}--\ref{FigC2} in Appendix \ref{appC}). At a conceptual level, the performance of any model-averaging framework is bounded by the expressive power of its constituent models: if the true epidemic dynamics deviate substantially from both the SEIR and Hawkes-type assumptions, the ensemble remains misspecified regardless of how weights are allocated. Addressing these limitations may require expanding the model set or incorporating diagnostics that detect systematic model inadequacy. Parameter estimation in each model was carried out using SMC$^2$, a method that may suffer from particle degeneracy when abrupt changes occur in the system (see SEIRS parameters in Figure \ref{FigC1} in Appendix \ref{appC}). Extensions to SMC proposed by \citet{birrell2020efficient}, designed to handle what they term “system shocks,” could offer a promising avenue for improving particle diversity and robustness under sudden regime changes.

Our study assumes that all cases arise from local transmission and are accurately reported. In practice, real outbreaks often involve imported infections, under-reporting, and reporting delays. Future work could extend the framework to explicitly account for these features, which would likely improve both realism and predictive reliability. Another promising direction is enhancing the DTHP model with more flexible triggering kernels. Current parametric kernels may miss complex transmission patterns; Bayesian spline-based kernels could provide a more adaptable, data-driven representation of self-excitation. 

In summary, our model-averaging framework leverages the complementary strengths of DTHP and SEIR(S) models to improve the robustness and reliability of epidemic inference and short-term forecasting. This highlights the framework’s value for adaptive inference while also emphasizing the need for methodological advances to maintain performance under sudden shifts in epidemic dynamics

\section*{Acknowledgments}
This publication has emanated from research conducted with the financial support of Taighde
Éireann – Research Ireland under Grant number 21/FFP-P/10123. 

\section*{Data availability}
All code used to produce the analyses in this paper is available at  
\href{https://github.com/Dhorasso/bma-smc2-dthp-seir}{https://github.com/Dhorasso/bma-smc2-dthp-seir}.  
The datasets used are publicly available from the HPSC at  
\href{https://respiratorydisease-hpscireland.hub.arcgis.com/pages/influenza}{https://respiratorydisease-hpscireland.hub.arcgis.com/pages/influenza}  
for influenza data, and from the Irish COVID-19 Data Hub at  
\href{https://COVID19ireland-geohive.hub.arcgis.com/}{https://COVID19ireland-geohive.hub.arcgis.com/}  
for COVID-19 data.

\appendix
\renewcommand{\thefigure}{\thesection\arabic{figure}}
\renewcommand{\thetable}{\thesection\arabic{table}}
\renewcommand{\thealgorithm}{\thesection\arabic{algorithm}}
\counterwithin{figure}{section}
\counterwithin{table}{section}
\section{State Forecasting}\label{appA}

Once we obtain the filtering estimate of the state vector up to the final time point $T$ under model $\mathcal{M}_k$, represented by the posterior $p(x_{k,T} | y_{1:T}, \mathcal{M}_k)$ for $k = 1,\ldots,K$, we can compute the $h$-step-ahead predictive density for $1 \leq h \leq H$, denoted by $p(x_{k,T+h} | y_{1:T}, \mathcal{M}_k)$. These predictive states are generated by propagating the state model forward without incorporating new observations. The model-specific $h$-step-ahead predictive distribution for $x_{k,T+h}$ is obtained by marginalizing over the parameters:
\begin{align}\label{marg_predictive_forecast}
p(x_{k,T+h} | y_{1:T}, \mathcal{M}_k)
&= \int p(x_{k,T+h} | y_{1:T}, \theta_k, \mathcal{M}_k)\,
p(\theta_k | y_{1:T}, \mathcal{M}_k)\, d\theta_k \notag\\
&= \int 
\left[
\prod_{t=T+1}^{T+h} p(x_{k,t} | x_{k,t-1}, \theta_k, \mathcal{M}_k)
\right]
p(x_{k,T} | y_{1:T}, \theta_k, \mathcal{M}_k)\,
p(\theta_k | y_{1:T}, \mathcal{M}_k)\,
dx_{_{k.T+1:T+h-1}}\, d\theta_k.
\end{align}
In practice, to make this computation tractable, we use a subset of $N_c$ parameter particles $\{\theta_k^j\}_{j=1}^{N_c}$ drawn from the posterior samples obtained via SMC$^2$ (see Algorithm~\ref{BMA-SMC2}). For each parameter particle $\theta_k^j$, we propagate $N_x$ state particles $\{x^i_{k,T+h,j}\}_{i=1}^{N_x}$ forward using the transition distribution $p(x_{k,T+h}| x_{k,T+h-1}, \theta_k^j, \mathcal{M}_k)$, and resample them based on the normalized state particle weights $W^{i}_{k,T,j}$ obtained at time $T$, as detailed in Algorithm~\ref{forecast}.

Bayesian model averaging combines the model-specific predictive distributions using the posterior model probabilities $\pi_{k,T}$, which are assumed constant over the forecast horizon:
\begin{align}
p(x_{T+h} | y_{1:T}) = \sum_{k=1}^{K} \pi_{k,T} \, p(x_{k,T+h} | y_{1:T}, \mathcal{M}_k),
\end{align}
This approach treats future observations as unobserved by design. Projecting the hidden states forward effectively generates samples $x_{k,T+h}$ for $h = 1,\dots,H$, and corresponding predictive distributions for observable quantities $y_{T+h}$ can then be obtained by drawing from the observation model conditional on these projected states. In this way, both latent and observed future quantities are forecasted consistently within the state-space framework. The model-averaged predictive mean for a shared quantity $\Delta_{T+h}$ is given by
\begin{align}
\mathrm{E}[\Delta_{T+h}|y_{1:T}] \approx \widehat{\Delta}_{T+h} = \sum_{k=1}^{K} \pi_{k,T} \, \widehat{\Delta}_{k,T+h}, \quad \text{for}\quad h=1,\dots,H,
\end{align}
where
\begin{align}
\mathrm{E}[\Delta_{k, T+h}|y_{1:T}] \approx \widehat{\Delta}_{k, T+h} =\dfrac{1}{N_cN_x}  \sum_{j=1}^{N_c} \sum_{i=1}^{N_x} \Delta^i_{k,T+h,j}.
\end{align}
\begin{algorithm}[H]
\caption{State Forecasting for model $\mathcal{M}_k$ with resampling}
\begin{algorithmic}[1]
\For{$h = 1$ to $H$} \Comment{forecast horizon}
    \For{$j = 1$ to $N_c$} \Comment{parameter particles}
        \For{$i = 1$ to $N_x$} \Comment{state particles}
            \State Sample $x^i_{k,T+h,j} \sim p(x_{k,T+h} | x^i_{k,T+h-1,j}, \theta^j_k, \mathcal{M}_k)$
        \EndFor
        \State Resample $\{x^i_{k,T+h,j}\}_{i=1}^{N_x}$ according to $\{W^i_{k,T,j}\}_{i=1}^{N_x}$
    \EndFor
    \State Compute model-specific predictive mean:
$\widehat{x}_{k,T+h} =\frac{1}{N_cN_x}  \sum_{j=1}^{N_c} \sum_{i=1}^{N_x} x^i_{k,T+h,j}$
\EndFor
\end{algorithmic}
\label{forecast}
\end{algorithm}

\section{Parameter filtered distribution and model weights in simulated data}\label{appB}

Preliminary experiments suggest that the model is generally not highly sensitive to most parameter settings, with the exception of $\gamma$ in the SEIR model, which is strongly correlated with the transmission rate $\beta_t$. This correlation is especially prominent during abrupt changes in transmission dynamics.  The priors on $\beta_{0}$ and $R_{0}$ are guided by the values used to generate the data for the different scenarios. The prior for $\nu_k$ allows for rapid changes in the transmission rate, ensuring the model adapts effectively to sudden shifts in disease dynamics, where higher values of $\nu_k$ correspond to larger changes in the transmission rate. 

\begin{figure}[H]
\centering
\includegraphics[width=0.88\linewidth]{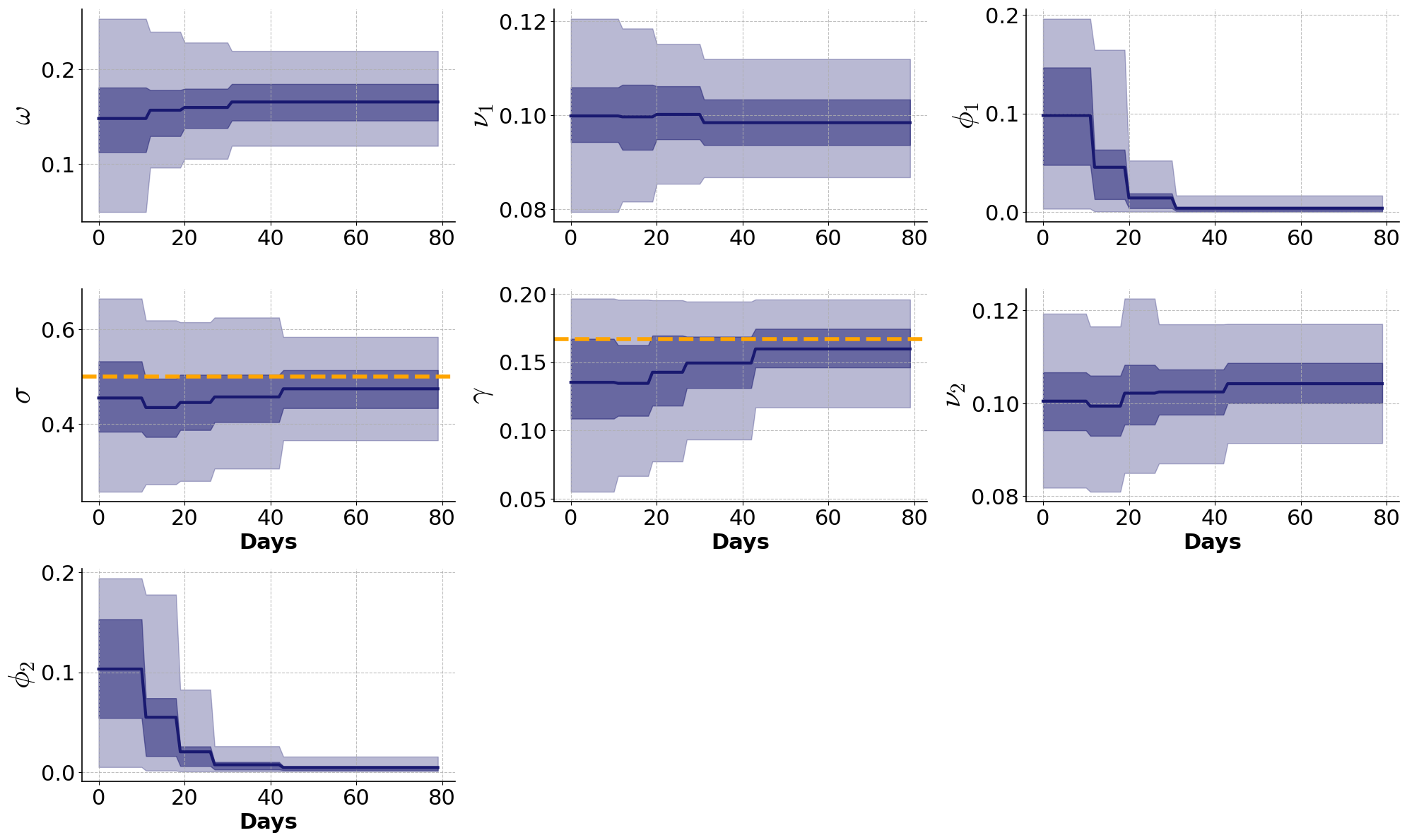}
\caption{\footnotesize Evolution of the posterior distributions of each parameter for the DTHP and SEIR models under Scenario A. Posterior means are shown as dark violet lines, with the darker and lighter violet areas indicating the 50\% and 95\% credible intervals, respectively. The true value of $\sigma$ and $\gamma$ used to generate the data is indicated by the horizontal orange dashed line.}
\label{FigB1}
\end{figure}

\begin{figure}[H]
\centering
\includegraphics[width=0.88\linewidth]{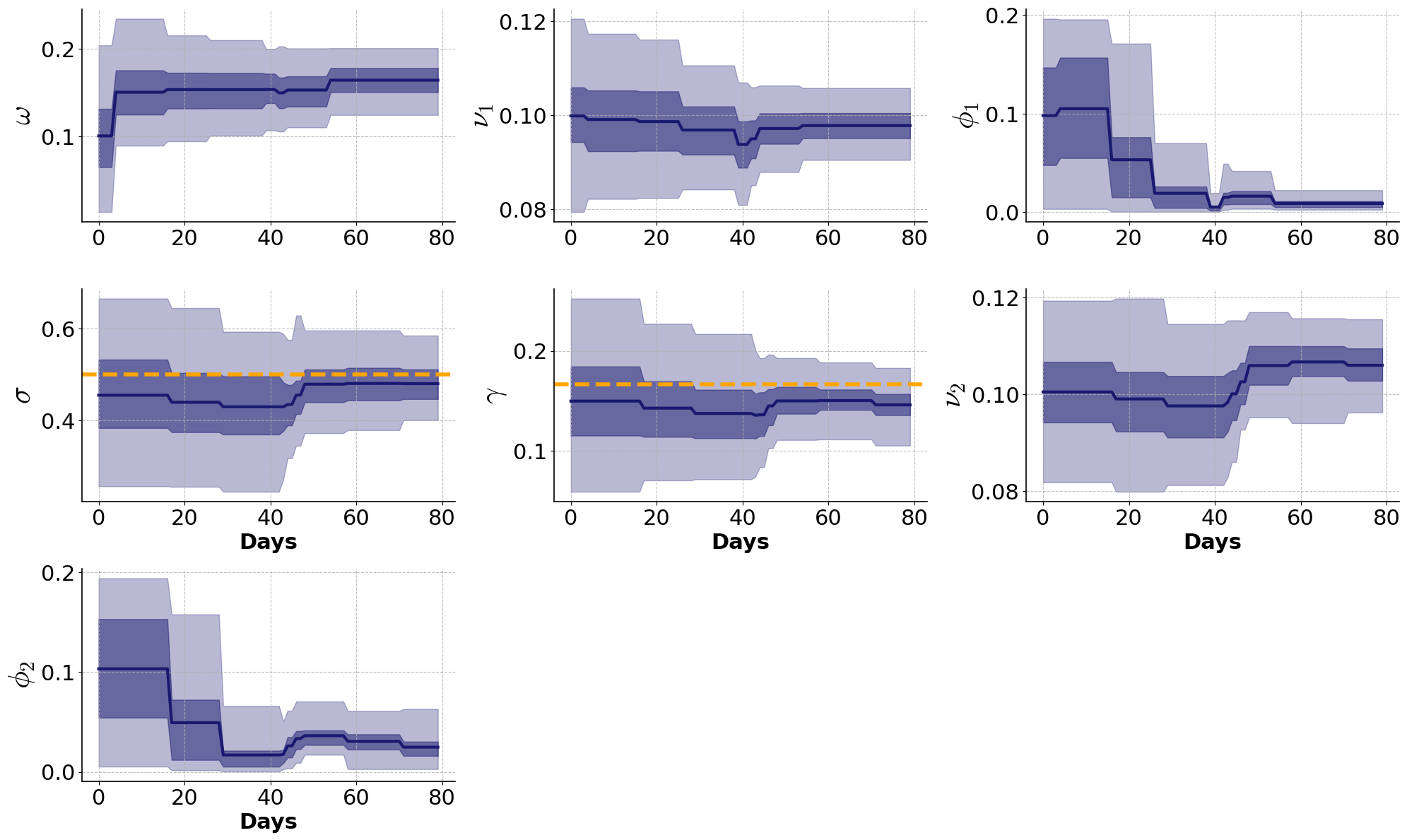}
\caption{\footnotesize Evolution of the posterior distributions of each parameter for the DTHP and SEIR models under Scenario B. Posterior means are shown as dark violet lines, with the darker and lighter violet areas indicating the 50\% and 95\% credible intervals, respectively. The true value of $\sigma$ and $\gamma$ used to generate the data is indicated by the horizontal orange dashed line.}
\label{FigB2}
\end{figure}

\begin{figure}[H]
\centering
\includegraphics[width=0.88\linewidth]{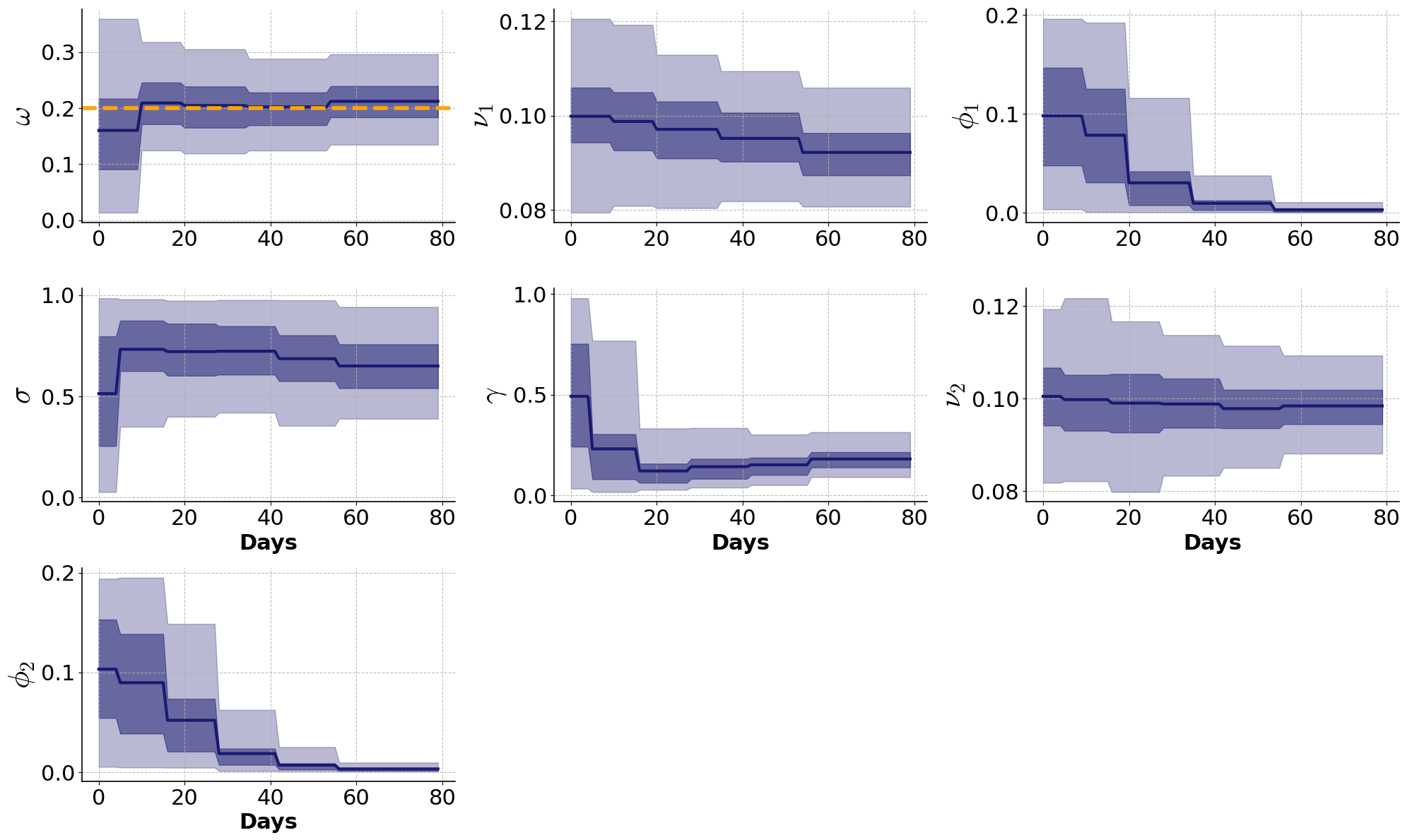}
\caption{\footnotesize Evolution of the posterior distributions of each parameter for the DTHP and SEIR models under Scenario C. Posterior means are shown as dark violet lines,  with the darker and lighter violet areas indicating the 50\% and 95\% credible intervals, respectively.  The true value of $\omega$ used to generate the data is indicated by the horizontal orange dashed line.}
\label{FigB3}
\end{figure}

\begin{figure}[H]
\centering
\includegraphics[width=0.7\linewidth]{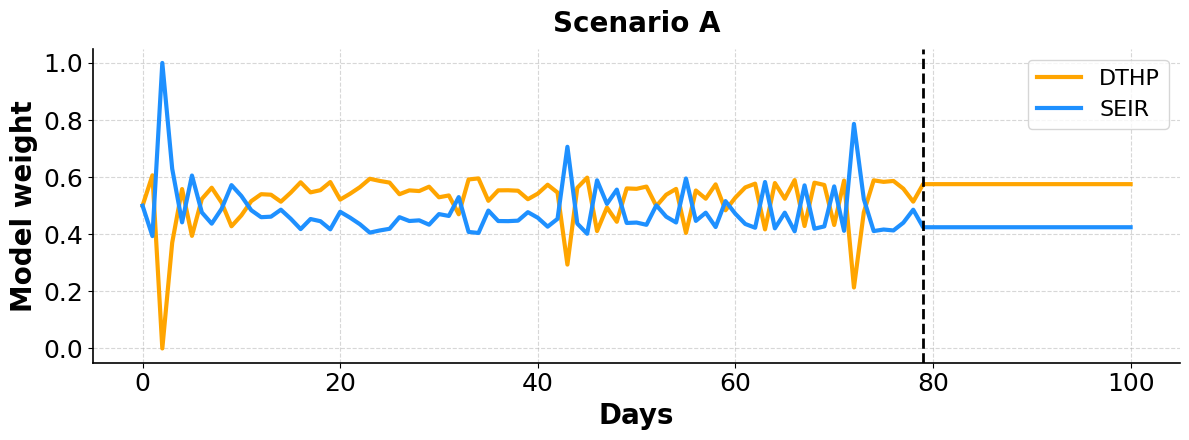}

\includegraphics[width=0.7\linewidth]{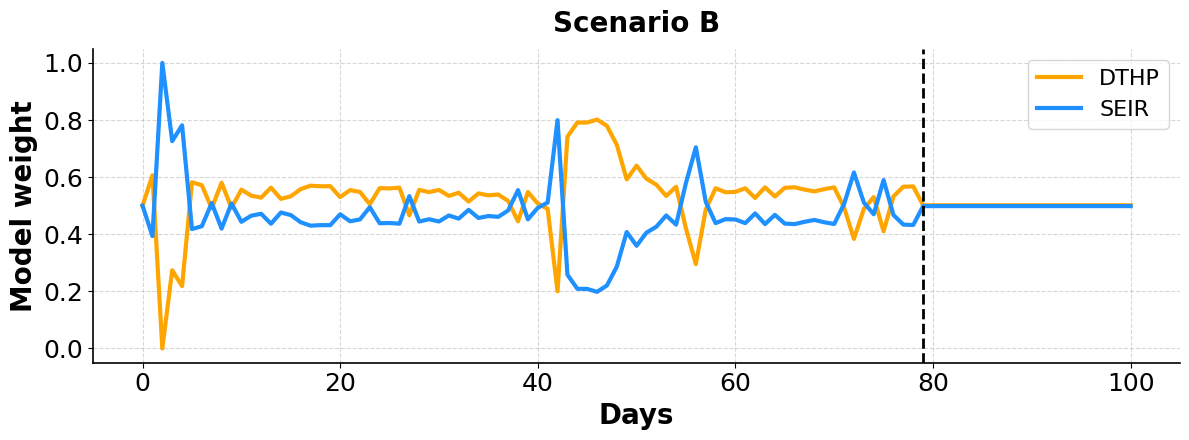}

\includegraphics[width=0.7\linewidth]{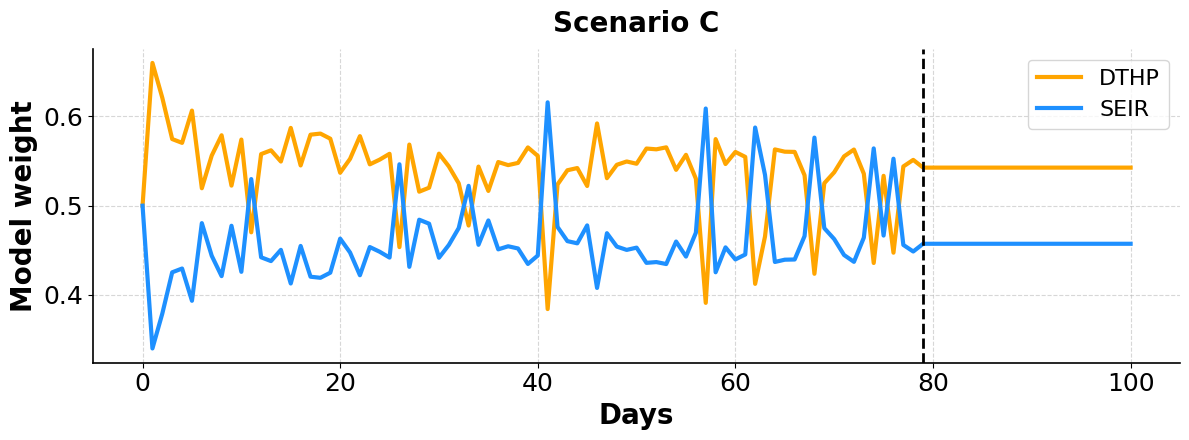}

\caption{\footnotesize Evolution of model weights across scenarios A, B, and C. The plot illustrates the evolution of model weights of the DTHP and the SEIR model. The vertical black dashed line indicates the start of the forecasting period.}
\label{FigB4}
\end{figure}

\section{SEIRS model Influenza and Prior distributions}\label{appC}
Given that the dataset spans a period of approximately two years, it is essential to incorporate demographic factors into our model and consider immunity loss. To account for these dynamics, we consider the following SEIRS model:

\begin{align}\label{seirs_model}
S(t+1) &= S(t) + \mu \big(N(t) - S(t)\big) - \lambda_{SE}(t) + \lambda_{RS}(t), \notag \\
E(t+1) &= E(t) + \lambda_{SE}(t) - \lambda_{EI}(t) - \mu E(t), \notag \\
I(t+1) &= I(t) + \lambda_{EI}(t) - \lambda_{IR}(t) - \mu I(t), \\
R(t+1) &= R(t) + \lambda_{IR}(t) - \mu R(t) - \lambda_{RS}(t). \notag
\end{align}
Here, the transitions between compartments are modeled as:
\begin{align}
   &\lambda_{SE}(t) \sim \text{Binomial}\left(S(t), 1 - e^{-\beta_t \frac{I(t)}{N(t)}}\right), \quad
\lambda_{EI}(t) \sim \text{Binomial}\left(E(t), 1 - e^{-\sigma}\right), \notag\\
&\lambda_{IR}(t) \sim \text{Binomial}\left(I(t), 1 - e^{-\gamma}\right), \quad \lambda_{RS}(t) \sim \text{Binomial}\left(R(t), 1 - e^{-\alpha}\right). 
\end{align}

The transmission rate $\beta_t$ as described in the main manuscript. The parameter $1/\alpha$ denotes the average duration of immunity, while $1/\sigma$ and $1/\gamma$ represent the average durations of the latent and infectious periods, respectively. The parameter $\mu$ denotes the recruitment or mortality rate.

For the SEIRS model, the initial state was set as $S(0) = N -E(0)- I(0)$, with $E(0) \sim \mathcal{U}(\{0, \dots, 5\})$, $I(0) \sim \mathcal{U}(\{0, \dots, 15\})$ and $R(0) = 0$, where the total initial population was fixed at $N = 5.16 \times 10^6$. The prior distributions for the parameters were specified as follows: $\mu = 1/(80 \times 52)$ weeks$^{-1}$, $\sigma \sim \mathcal{TN}_{[7/3,~7]}(7/1.5,~0.1^2)$,  $\gamma \sim \mathcal{TN}_{[1,~7/5]}(7/6,~0.1^2)$, $\alpha \sim \mathcal{U}([1/48,~1/24])$, and $\beta_{0} \sim \mathcal{TN}_{[0, 1]}(0.4,~0.05^2)$. For the DTHP model, the initial conditions were defined as $\lambda_{H}(0) \sim \mathcal{U}(\{0, \dots, 15\})$, $R_{0} \sim \mathcal{N}(0.5,~0.05^2)$, and $\omega \sim \mathcal{U}([0,~1])$. The hyperparameters were chosen as $\nu_k \sim  \mathcal{TN}_{[0.05, 0.15]}(0.1,~0.02^2)$ and $\phi_k \sim \mathcal{U}([0,~0.2])$ for $k = 1, 2$.

\begin{figure}[H]

    \centering
    \includegraphics[width=1\linewidth]{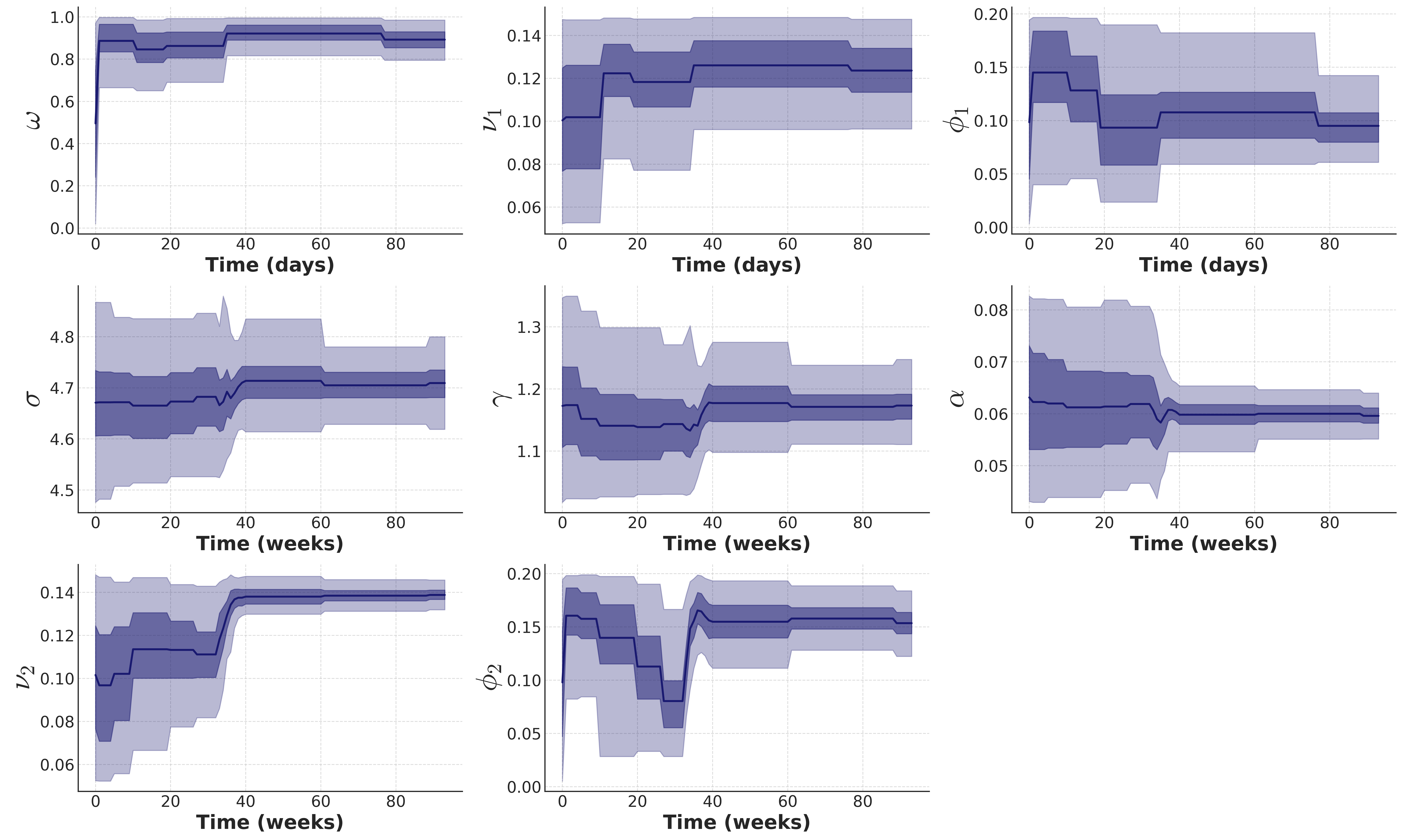}
  \caption{\footnotesize Evolution of the weekly estimate of the posterior distributions of each parameter for the DTHP and SEIRS models obtained using SMC$^2$ for Influenza data. Model posterior means are shown as dark violet lines, with the violet band representing the 95\% credible intervals.}
    \label{FigC1}
\end{figure}

For COVID-19, we used the SEIR model with initial conditions specified as $S(0) = N - E(0) - I(0)$, $E(0) = 5$, $I(0) \sim \mathcal{U}(\{0, \dots, 15\})$, and $R(0) = 0$. The prior distributions were defined as $\beta_0 \sim \mathcal{N}(0.5,~0.05^2)$, $\sigma \sim \mathcal{TN}_{[1/5,~1/3]}(1/4,~0.1^2)$, and $\gamma \sim \mathcal{TN}_{[1/7.5,~1/4.5]}(1/6,~0.2^2)$. For the DTHP model, the initial conditions were $\lambda_{H}(0) \sim \mathcal{U}(\{0, \dots,15\})$, $R_{0} \sim \mathcal{N}(3.2,~0.05^2)$, and $\omega \sim \mathcal{U}([0,~1])$. The hyperparameters were set as $\nu_k \sim  \mathcal{TN}_{[0.05, 0.15]}(0.1,~0.02^2)$ and $\phi_k \sim \mathcal{U}([0,~0.2])$ for $k = 1, 2$. 

\begin{figure}[H]

    \centering
    \includegraphics[width=1\linewidth]{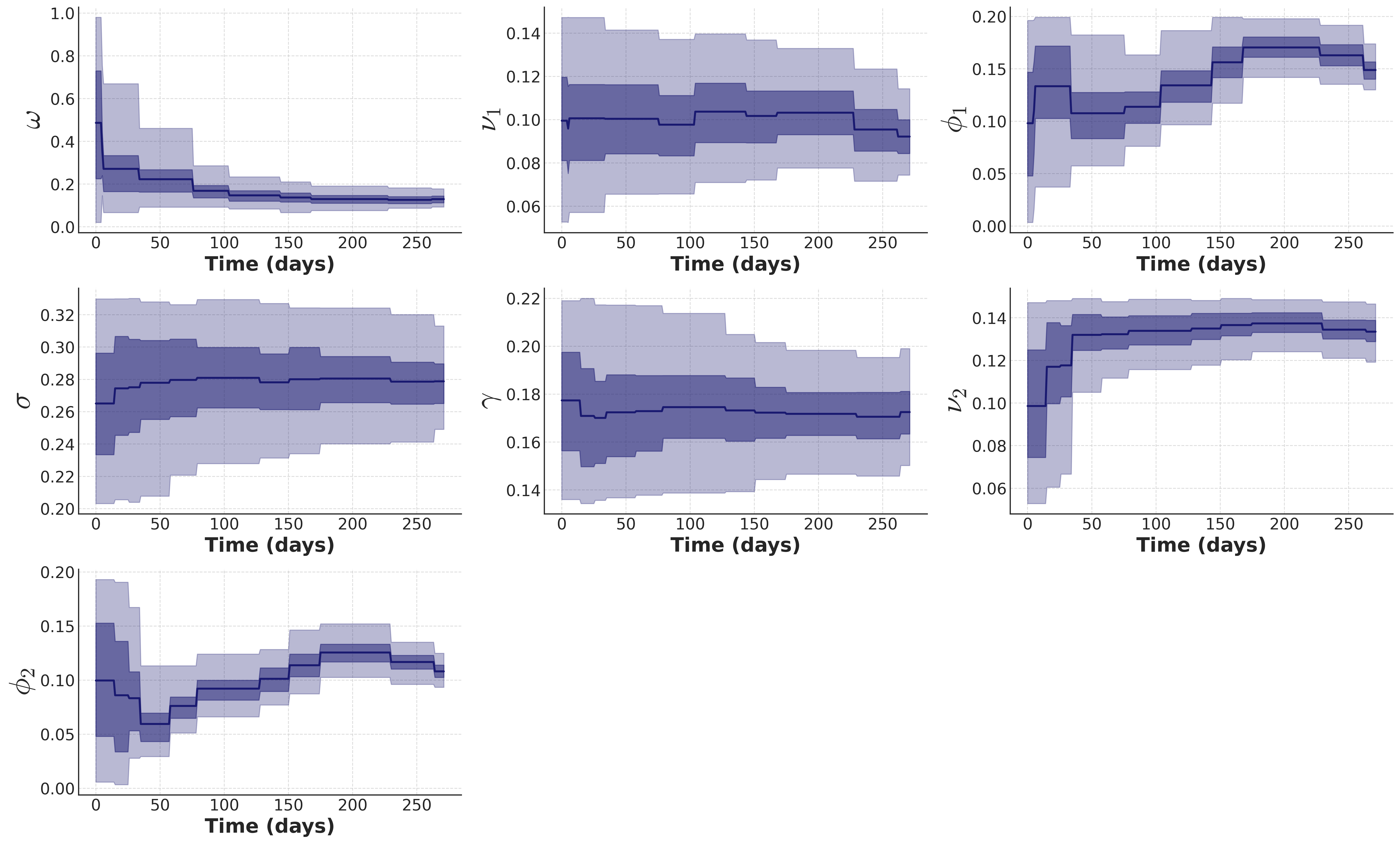}
  \caption{\footnotesize Evolution of the daily estimate of the posterior distributions of each parameter for the DTHP and SEIR models obtained using SMC$^2$ for COVID-19 data. Model posterior means are shown as dark violet lines, with the violet band representing the 95\% credible intervals.}
    \label{FigC2}
\end{figure}

\begin{figure}[H]
    \centering
    \includegraphics[width=0.7\linewidth]{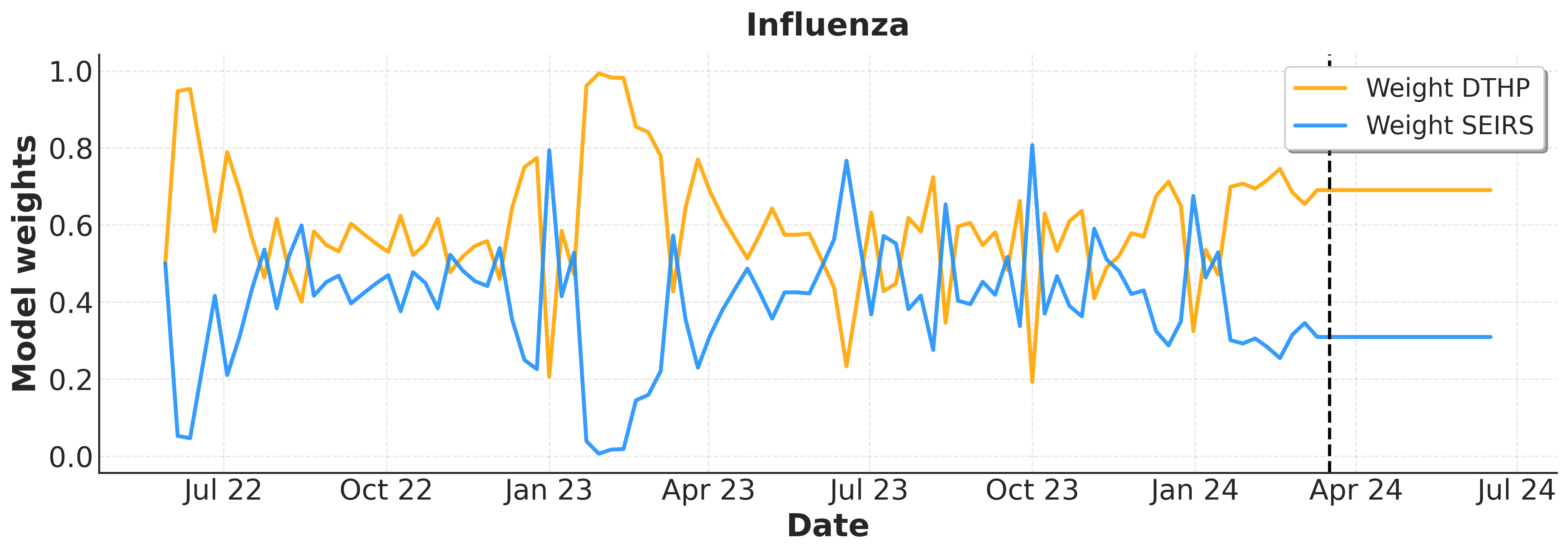}
 \includegraphics[width=0.7\linewidth]{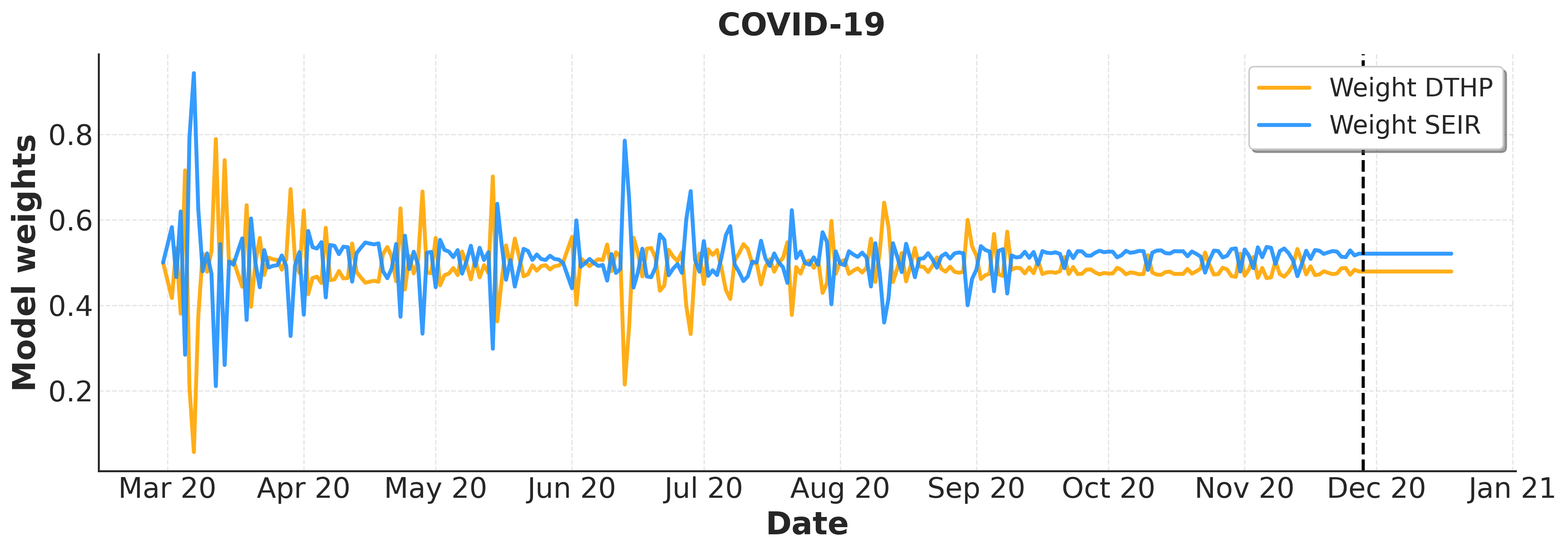}
    \caption{\footnotesize Evolution of model weights for the Influenza and  COVID-19 Dataset. The vertical black dashed line indicates the start of the forecasting period.}
    \label{FigC3}
\end{figure}

\bibliographystyle{agsm}
\bibliography{references}

\end{document}